\documentclass[journal,comsoc]{IEEEtran}

\usepackage{amsmath,amsfonts}
\usepackage{algorithmic}
\usepackage[linesnumbered,ruled,vlined]{algorithm2e}
\usepackage{array}
\usepackage{textcomp}
\usepackage{stfloats}
\usepackage{url}
\usepackage{verbatim}
\usepackage{graphicx}
\usepackage{cite}
\usepackage{multirow}
\usepackage{makecell}
\usepackage{pifont}
\usepackage{tikz}
\newcommand*\circled[1]{\tikz[baseline=(char.base)]{
		\node[shape=circle,draw,inner sep=1.2pt] (char) {#1};}}
\usepackage{enumitem}
\usepackage{subcaption} 
\usepackage{caption}
\allowdisplaybreaks
\newtheorem{assumption}{\textbf{Assumption}}
\newtheorem{theorem}{\textbf{Theorem}}

\newtheorem{lemma}{\textbf{Lemma}}
\newtheorem{corollary}{\textbf{Corollary}}
\newtheorem{remark}{\textbf{Remark}}

\newtheorem{observation}{\textbf{Observation}}

\begin{document}

\title{Federated Attention: A Distributed Paradigm for Collaborative LLM Inference over Edge Networks}

\author{Xiumei Deng, Zehui Xiong, Binbin Chen, Dong In Kim, Merouane Debbah, H. Vincent Poor
        % <-this % stops a space}
        
\thanks{X. Deng and B. Chen are with the Pillar of Information Systems Technology and Design, Singapore University of Technology and Design, Singapore (e-mail: \{xiumei\_deng, binbin\_chen\}@sutd.edu.sg).}
\thanks{Z. Xiong is with the School of Electronics, Electrical Engineering and Computer Science, Queen's University Belfast, Belfast, BT7 1NN, U.K. (e-mail: z.xiong@qub.ac.uk).}
\thanks{D. I. Kim is with the Department of Electrical and Computer Engineering, Sungkyunkwan University, Suwon 16419, South Korea (e-mail: dongin@skku.edu).}
\thanks{M. Debbah is with the KU 6G Research Center, Department of Computer and Information Engineering, Khalifa University, Abu Dhabi, UAE, and also with the CentraleSupelec, University of Paris-Saclay, 91192 Gif-surYvette, France (e-mail: merouane.debbah@ku.ac.ae).}
\thanks{H. V. Poor is with Department of Electrical and Computer Engineering, Princeton University, NJ 08544, USA (e-mail: poor@princeton.edu).}
%\thanks{Manuscript received April 19, 2021; revised August 16, 2021.}
}
\markboth{}%
{Shell \MakeLowercase{\textit{et al.}}: A Sample Article Using IEEEtran.cls for IEEE Journals}


\markboth{This work has been submitted to the IEEE Journal on Selected Areas in Communications (JSAC) for possible publication.
}%
{How to Use the IEEEtran \LaTeX \ Templates}

\maketitle

\begin{abstract}
	Large language models (LLMs) are proliferating rapidly at the edge, delivering intelligent capabilities across diverse application scenarios such as Industrial Internet of Things (IIoT), intelligent transportation systems (ITS), and smart home. However, their practical deployment in collaborative scenarios confronts fundamental challenges: privacy vulnerabilities, communication overhead, and computational bottlenecks. To address these challenges, we propose Federated Attention (FedAttn), which integrates the federated paradigm into the self-attention mechanism, creating a new distributed LLM inference framework that simultaneously achieves privacy protection, communication efficiency, and computational efficiency. Specifically, FedAttn enables participants to perform local self-attention over their own token representations while periodically exchanging and aggregating Key-Value (KV) matrices across multiple Transformer blocks, collaboratively generating LLM responses without exposing private prompts. Further, we identify a structural duality between contextual representation refinement in FedAttn and parameter optimization in federated learning across three pillars: private data, local computation, and global aggregation. This key insight provides a principled foundation for systematically porting federated optimization techniques to collaborative LLM inference. Building on this framework, we theoretically analyze how local self-attention computation within participants and heterogeneous token relevance among participants shape error propagation dynamics across Transformer blocks. Moreover, we characterize the fundamental trade-off between response quality and communication/computation efficiency, which is governed by the synchronization interval and the number of participants. The experimental results validate our theoretical analysis, and reveal significant optimization opportunities through sparse attention and adaptive KV aggregation, highlighting FedAttn's strong potential to deliver scalability and efficiency in real-world edge deployments.
\end{abstract}


\section{Introduction}
\IEEEPARstart{T}{he} year 2022 marked a pivotal moment in the evolution of generative artificial intelligence (GAI) when OpenAI's release of ChatGPT triggered unprecedented global interest, fundamentally redefining AI's potential to transform industries and human activities. Propelled by intense competition among leading technology corporations developing proprietary foundation models, large language models (LLMs) have rapidly evolved to handle increasingly complex tasks through innovations such as Mixture of Experts (MoE) and Chain of Thought (CoT) prompting, demonstrating their remarkable potential for solving real-world problems\cite{DBLP:journals/tai/HagosBR24,DBLP:conf/icaiic/Saha24}.

However, these technological advances come with significant computational complexity, most notably stemming from the Transformer's self-attention mechanism that exhibits linear scaling with model size and quadratic scaling with sequence length. This fundamental challenge is further exacerbated as modern LLMs continue to advance along\textit{ 1) increasing model size}, and \textit{2) expanding input sequence lengths} required for long-context tasks such as document analysis and code generation, further amplified by prompt engineering techniques such as retrieval-augmented generation (RAG) for external knowledge integration and CoT for complex reasoning. Consequently, the escalating computational burden renders LLM inference prohibitively expensive, requiring substantial investment in hardware infrastructure and incurring significant energy costs, ultimately impeding the deployment of LLMs in real-world applications.

Currently, practical LLM deployment predominantly adopts two paradigms: \textit{1) cloud inference} and \textit{2) on-device inference}. Cloud LLM inference\cite{DBLP:conf/hpec/Li0GT24}, exemplified by ChatGPT, operates whereby user prompts are transmitted to remote servers for processing by high-performance computing infrastructure. This paradigm presents two critical challenges: \textit{1) privacy and security vulnerabilities}, including potential data disclosure and unauthorized access to sensitive and personally identifiable information, which are particularly problematic in data-sensitive sectors such as finance and legal services, as well as in safety-critical areas such as healthcare, where privacy breaches may violate data protection regulations such as the EU's General Data Protection Regulation (GDPR); and \textit{2) communication delays}, which are especially pronounced in wireless networks where heavy data traffic between end devices and a remote cloud can overwhelm the limited transmission bandwidth, and are particularly problematic in latency-sensitive applications such as autonomous vehicles. On-device LLM inference\cite{11177548} addresses these challenges by processing user prompts locally. However, this paradigm faces the critical \textit{3) computation bottleneck}: modern LLMs demand substantial memory and computing power that typically surpass the capabilities of user devices, rendering on-device inference infeasible.

Furthermore, the aforementioned challenges of both LLM inference paradigms are further exacerbated in collaborative scenarios\cite{DBLP:conf/infocom/MaGZ025,DBLP:journals/corr/abs-2507-16731}, involving multiple participants engaging in collective reasoning and decision-making. Since cloud LLM inference requires all participants to transmit their private prompts to remote servers before processing, it not only exposes every single participant to privacy and security vulnerabilities, but also prolongs communication delays due to synchronization barriers across participants. For on-device LLM inference, the extended input sequence length from multiple participants substantially increases computational complexity, straining the limited computational resources of user devices.

These fundamental challenges underscore the critical need for a collaborative LLM inference paradigm that achieves privacy protection, communication efficiency, and computational efficiency. Furthermore, this imperative has intensified with the proliferation of AI applications in edge networks across diverse domains including Industrial Internet of Things (IIoT) and intelligent transportation system (ITS), which inherently feature collaborative tasks among multiple participants operating under limited computational and communication resources\cite{10835069,DBLP:journals/csur/ZhengCQSSC25,DBLP:journals/twc/XieXZXGGP24}. For instance, autonomous driving systems require individual vehicles to share sensor data and trajectory predictions over mobile networks in order to collaboratively query LLMs for right-of-way negotiation in scenarios such as highway merging, thereby facilitating navigation with real-time collision avoidance and cooperative traffic flow optimization. Such emerging edge AI applications not only demand modern LLMs for processing complex queries, but also increasingly expect LLMs to deliver responses at the edge rather than in remote clouds to address the stringent requirements of both real-time responsiveness and data privacy.

To this end, we propose Federated Attention (FedAttn), a new distributed self-attention paradigm tailored for the non-autoregressive components of Transformers, enabling multiple participants to collaboratively generate LLM responses without sharing their private prompts. As dual implementations of federated paradigm for model inference and training, our proposed FedAttn and federated learning (FL) share the following core principles:\begin{enumerate}[label=\bf\textit{\arabic*)}]
	\item \textbf{{Privacy protection. }}Eliminating the need for raw data sharing via local computation and global aggregation.
	\item \textbf{{Computation efficiency. }}Reducing computational and memory complexities via distributed parallel computing.	 
	\item \textbf{{Communication efficiency.}} Minimizing overall communication overhead via periodic synchronization rounds.\end{enumerate}These key advantages can collectively empower collaborative LLM inference tasks under computational and communication resource constraints, paving the way for large-scale LLM deployment in practical edge networks. The key contributions of this paper are summarized as follows.\begin{enumerate}[label=\bf\textit{\arabic*)}]
	\item \textbf{Federated Attention}. The main idea of FedAttn is that participants execute self-attention mechanisms on their local token representations and periodically exchange local Key-Value (KV) matrices at intervals of a certain number of Transformer blocks, which are then aggregated into a global KV matrix that kickstarts each participant's local self-attention computation for subsequent intervals.
	\item \textbf{Federated Duality}. We formalize a structural duality between FedAttn and FL across three key dimensions of the federated paradigm: \textit{1) Private data:} FedAttn infers a collective response from individual user prompts, while FL trains a global model over local datasets. \textit{2) Local computation: }FedAttn refines local token representations through successive forward passes, each performing local self-attention over one Transformer block, while FL optimizes local model parameters through iterative backward passes, each performing one step of local gradient descent. \textit{3) Global aggregation:} FedAttn constrains attention to local KVs and expands to global contextual information every several blocks, mirroring FL's periodic model aggregation to learn global knowledge from local models. 
	\item \textbf{Error analysis}. We theoretically analyze the error propagation dynamics of FedAttn across Transformer blocks, which primarily reveals that {\textit{1) Approximation error}} increases monotonically with local forwards $H$, formally establishing the trade-off between response quality and communication efficiency. {\textit{2) Marginal communication benefit}} identifies small $H$ values as a critical regime where FedAttn achieves substantial communication savings with limited degradation in response quality. {\textit{3) Blocks in shallow layers}} dominate error accumulation, suggesting their prioritization for performing global self-attention to minimize overall approximation error.	
	\item \textbf{Experimental findings}. We conduct experiments to evaluate the efficacy and efficiency of FedAttn on Qwen2.5 models using GSM8K. Experimental results verify theoretical analysis, demonstrating {\textit{1) trade-offs}} between response quality and communication/computational cost, {\textit{2) error propagation dynamics}} across blocks, and {\textit{3) token relevance}} both within and across participants. Building upon these findings, we further investigate optimization opportunities by integrating sparse self-attention mechanisms into FedAttn: {\textit{1) Sparse local attention}} randomly samples input tokens before local computation, reducing computational cost at the expense of response quality. {\textit{2) Sparse KV exchange}} randomly samples local KVs before global aggregation, remarkably improving response quality while reducing communication cost. These experimental results demonstrate FedAttn's effectiveness and robustness to computational and communication resource limitations, highlighting its practical viability for distributed LLM inference in edge networks.\end{enumerate}
	
The remainder of this paper is organized as follows. Section \ref{Related work} presents related works and significance of our work. Section \ref{Revisiting Self-Attention Mechanism} provides some background on self-Attention mechanism. Section \ref{Federated Attention Paradigm} presents our proposed FedAttn paradigm. Section \ref{Federated Duality: From FL to FedAttn} demonstrates the duality between FedAttn and FL, followed by Section \ref{Error Analysis} that theoretically analyzes the error propagation of FedAttn. Section \ref{Experimental Results} represents the experimental results. Section \ref{Conclusion} concludes this paper.

\section{Related Work and Significance}\label{Related work}

This section reviews existing distributed LLM inference frameworks and their underlying model parallel computing paradigms, analyzing their fundamental limitations. We then clarify how our federated attention paradigm addresses these research gaps, and elucidates its significance.

\subsection{Distributed LLM Inference}
Cloud LLM inference exposes user prompts to remote servers, raising critical privacy and security concerns while introducing substantial communication delays, whereas on-device inference is constrained by insufficient computational resources to execute modern LLMs. Recent research has extensively explored distributed computing as a promising alternative for deploying LLMs in practical applications. This approach involves distributing LLM inference workloads across multiple end devices and edge servers, in order to \textit{1) alleviate the substantial computational demands} of advanced LLMs, \textit{2) mitigate privacy and security risks} by preventing any individual node from accessing the complete user prompts, and \textit{3) reduce communication overhead} by utilizing edge infrastructure geographically closer to users than remote clouds.

Existing studies on distributed LLM inference primarily concentrate on computation offloading scenarios, where users distribute their inference workloads across multiple end devices and edge nodes. Substantial effort has been devoted to optimizing algorithms and allocating resources to enhance computational and communication efficiency of LLM inference in edge networks. For instance, \cite{DBLP:journals/iotj/ZhangSCCJ25} and \cite{DBLP:conf/lanman/MacarioSK25} propose distributed LLM inference frameworks wherein users locally execute tokenizer and embedding layers to process private prompts, then offload subsequent Transformer blocks to multiple edge servers. In a sequential manner, each server executes its assigned segment of consecutive Transformer blocks by accepting intermediate activations from the previous server and forwarding outputs to the next server, ultimately producing the final LLM response. Several works \cite{10681712,DBLP:conf/icc/FengLLCZZTG25,DBLP:journals/corr/abs-2502-12574} propose to partition self-attention computations across multiple attention heads, and decompose feed-forward networks (FFNs) through partitioning the first linear transformation into columns and the second into rows, thereby enabling distributed LLM inference in a parallel computing fashion. 

Alternative studies propose to segment user prompts during LLM inference. For instance, \cite{DBLP:conf/icml/MaiYHYP24} decomposes user prompts into sub-prompts and pseudo-prompts to obfuscate sensitive information across multiple devices for privacy preservation. \cite{DBLP:conf/mlsys/GimCLSK024} modularizes prompt segments into reusable modules within trusted nodes for minimizing redundant computations, thereby reducing the overall inference latency. Extending this approach of prompt segmentation, distributed MoE frameworks \cite{DBLP:journals/corr/abs-2505-13345,DBLP:conf/usenix/LiJZ0X23,DBLP:journals/corr/abs-2509-25041,DBLP:journals/corr/abs-2508-12851,DBLP:journals/corr/abs-2508-09208} further advance distributed LLM inference through routing individual input tokens to specialized expert modules across multiple nodes, producing final responses via parallel expert processing and output aggregation.

\subsection{Model Parallel Computing Paradigm}
Although prior studies have developed various distributed LLM inference frameworks to partition computational workloads, they fundamentally rely on existing model parallelism paradigms, which we review below:\begin{enumerate}[label=\bf\textit{\arabic*)}] 
	\item\textit{Pipeline parallelism} partitions the model into consecutive segments along layer dimension, where intermediate activations flow sequentially through these segments across computing nodes\cite{DBLP:journals/iotj/ZhangSCCJ25, DBLP:conf/lanman/MacarioSK25}. 
	\item\textit{Tensor parallelism} partitions the model along the hidden dimension, with each node computing a shard of matrix operations and executing all-reduce and all-gather operations to reconstruct the complete intermediate activations\cite{10681712,DBLP:conf/icc/FengLLCZZTG25,DBLP:journals/corr/abs-2502-12574}.
	\item\textit{Expert parallelism} distributes MoE experts across nodes, with a gating network dynamically routing input tokens to specialized experts\cite{DBLP:journals/corr/abs-2505-13345,DBLP:conf/usenix/LiJZ0X23,DBLP:journals/corr/abs-2509-25041,DBLP:journals/corr/abs-2508-12851,DBLP:journals/corr/abs-2508-09208}.\end{enumerate} 

Despite their widespread adoption in existing studies, these conventional paradigms face critical limitations when it comes to distributed LLM inference, as follows.
\begin{enumerate}[label=\bf\textit{\arabic*)}]
\item \textit{Communication Costs. }In pipeline parallelism, each model segment transmits hidden representations to the subsequent segment, incurring substantial communication cost that grows with pipeline depth, hidden dimension size, and sequence length. Tensor parallelism suffers from more severe communication bottlenecks due to the massive data transmission volume required for frequent all-reduce and all-gather operations after each linear transformation within self-attention mechanisms and FFNs, scaling linearly with hidden dimension and sequence length. To mitigate these communication overheads, recent works\cite{DBLP:journals/corr/abs-2503-01704,DBLP:conf/mlsys/0002L0XV24,DBLP:journals/corr/abs-2412-04964,DBLP:journals/corr/abs-2411-07942,DBLP:journals/corr/abs-2411-09510,DBLP:journals/corr/abs-2411-02829} propose applying sparsification and quantization to compress intermediate activations before data transmission. Considering heterogeneous communication and computation resources across multiple devices and edge servers, other works \cite{DBLP:journals/corr/abs-2503-14882,DBLP:conf/wiopt/KafetzisKK25,DBLP:journals/corr/abs-2501-14205,zhao-etal-2024-lingualinked,DBLP:journals/corr/abs-2507-21276,DBLP:conf/iwqos/ZhuZXD25} develop optimization frameworks that jointly determine device selection, model partitioning, task offloading, and resource allocation (e.g., computational frequency, transmission power, bandwidth) to minimize inference latency or energy consumption. 

\item \textit{Varying Sequence Lengths.} Despite optimization efforts\cite{DBLP:journals/iotj/ZhangSCCJ25, DBLP:conf/lanman/MacarioSK25, 10681712,DBLP:conf/icc/FengLLCZZTG25,DBLP:journals/corr/abs-2502-12574,DBLP:journals/corr/abs-2503-01704,DBLP:conf/mlsys/0002L0XV24,DBLP:journals/corr/abs-2412-04964,DBLP:journals/corr/abs-2411-07942,DBLP:journals/corr/abs-2411-09510,DBLP:journals/corr/abs-2411-02829,DBLP:journals/corr/abs-2503-14882,DBLP:conf/wiopt/KafetzisKK25,DBLP:journals/corr/abs-2501-14205,zhao-etal-2024-lingualinked,DBLP:journals/corr/abs-2507-21276,DBLP:conf/iwqos/ZhuZXD25,DBLP:journals/corr/abs-2505-13345,DBLP:conf/usenix/LiJZ0X23,DBLP:journals/corr/abs-2509-25041,DBLP:journals/corr/abs-2508-12851,DBLP:journals/corr/abs-2508-09208,DBLP:conf/icml/MaiYHYP24,DBLP:conf/mlsys/GimCLSK024}, pipeline and tensor parallelism paradigms face fundamental challenges when applied to distributed LLM inference. Unlike conventional AI models with fixed input and output dimensions, Transformers process input sequences of arbitrary length and generate outputs of uncertain length, exhibiting quadratic computational complexity along with unpredictable workloads that varies significantly across tasks. Studies building upon pipeline and tensor parallelism process each complete sequence within individual nodes, failing to adapt to task-varying computational demands. Consequently, exceptionally long sequences trigger memory overflow or severe load imbalances, requiring workload reallocation and data migration across nodes, degrading overall inference efficiency. 

\item \textit{Privacy Vulnerabilities }persist across the existing distributed LLM inference frameworks. Pipeline parallelism transmits hidden representations between nodes, which are essentially high-dimensional encodings of user prompts, and sensitive information leakage escalates proportionally to the pipeline depth. Expert parallelism suffers more severe privacy vulnerabilities\cite{DBLP:journals/corr/abs-2505-13345,DBLP:conf/usenix/LiJZ0X23,DBLP:journals/corr/abs-2509-25041,DBLP:journals/corr/abs-2508-12851,DBLP:journals/corr/abs-2508-09208}, as gating networks must access complete input sequences before routing that creates a single point of privacy risk, not to mention that tokens are shared directly with distributed experts across nodes.
\end{enumerate}

Most critically, existing frameworks\cite{DBLP:journals/iotj/ZhangSCCJ25, DBLP:conf/lanman/MacarioSK25, 10681712,DBLP:conf/icc/FengLLCZZTG25,DBLP:journals/corr/abs-2502-12574,DBLP:journals/corr/abs-2503-01704,DBLP:conf/mlsys/0002L0XV24,DBLP:journals/corr/abs-2412-04964,DBLP:journals/corr/abs-2411-07942,DBLP:journals/corr/abs-2411-09510,DBLP:journals/corr/abs-2411-02829,DBLP:journals/corr/abs-2503-14882,DBLP:conf/wiopt/KafetzisKK25,DBLP:journals/corr/abs-2501-14205,zhao-etal-2024-lingualinked,DBLP:journals/corr/abs-2507-21276,DBLP:conf/iwqos/ZhuZXD25,DBLP:journals/corr/abs-2505-13345,DBLP:conf/usenix/LiJZ0X23,DBLP:journals/corr/abs-2509-25041,DBLP:journals/corr/abs-2508-12851,DBLP:journals/corr/abs-2508-09208,DBLP:conf/icml/MaiYHYP24,DBLP:conf/mlsys/GimCLSK024} are fundamentally incompatible with collaborative scenarios where multiple participants each hold private prompts that collectively form the complete input of each inference task. These frameworks rely on the assumption that complete prompts resides on single nodes, focusing on computation offloading scenarios where individual users process their own prompts locally for their own inference tasks before offloading the remaining workloads to multiple nodes. However, collaborative scenarios present a distinct challenge: 

\textit{``Multiple users jointly perform LLM inference with each contributing prompt segments that collectively constitutes the complete input, yet none willing to reveal private prompts to each other."} 

This reveals a critical research gap, i.e., the absence of a privacy-preserving distributed LLM inference paradigm supporting collaborative scenarios.

\subsection{Research Gap and Significance}

To the best of our knowledge, FedAttn presents the first attempt to integrate the federated paradigm into self-attention mechanism at the core of Transformer-based LLMs, characterized by three key principles:\begin{enumerate}[label=\bf\textit{\arabic*)}]
	\item\textit{Privacy protection. }By enabling participants to perform local self-attention and exchange KVs rather than raw prompts, FedAttn substantially mitigates privacy vulnerabilities in collaborative LLM inference tasks. 
	\item\textit{Communication efficiency. }FedAttn reduces communication cost via periodic synchronization of KVs. Notably, modern LLMs are increasingly reducing the KV size using techniques such as Grouped Query Attention\cite{10_1145/3768165} that groups Queries to share KVs. FedAttn directly benefits from the reduced KV transmission volume, making it increasingly viable for practical deployment. 
	\item\textit{Computation efficiency. }Through distributing computational workload along the sequence dimension, FedAttn can easily accommodate heterogeneous device capabilities and adapt to varying computational complexity across tasks. \end{enumerate}In this way, FedAttn offers a new distributed LLM inference framework that achieves privacy preservation with communication and computation efficiency, making collaborative LLM applications practically viable over the resource-constrained edge networks.

Additionally, FedAttn exhibits key advantages compared to conventional model parallelism techniques. \textit{1) Unlike pipeline parallelism}, it enables parallel computing without the need for scheduling algorithms to reduce pipeline bubbles. \textit{2) Unlike tensor parallelism}, FedAttn substantially reduces communication overhead by avoiding the frequent all-reduce and all-gather operations. \textit{3) Unlike expert parallelism} limited to MoE, FedAttn is applicable to all Transformer-based LLMs.

Most crucially, we develop a structural duality between FedAttn and FL, revealing that FedAttn instantiates FL principles in the regime of collaborative LLM inference. This theoretical framework draws a parallel between federated model parameter optimization and collaborative contextual representation refinement, which provides a rigorous foundation for systematically transferring \textbf{FL Optimization Toolkits} to the emerging field of collaborative LLM inference as detailed below.
\begin{enumerate}[label=\bf\textit{\arabic*)}]
	\item \textit{From a theoretical perspective,} FL model convergence analysis of local training iterations and data heterogeneity directly translate to FedAttn, where we investigate how local attention computations and token relevance within and across participants, termed attention distribution in our work, impact LLM response quality. 

	\item \textit{From an algorithmic perspective}, FedAttn inherits extensive optimization strategies from FL to enhance efficacy and efficiency. \textit{a) Sparse attention} can reduce computational overhead while preserving response quality by applying techniques such as Sliding Window and Neighborhood Attention, whereby high-relevance tokens contribute more in local and global attention computations. \textit{b) Compressed KV exchange} using techniques such as sparsification and quantization can reduce communication overhead while preserving response quality. \textit{c) Adaptive KV aggregation} method prioritize critical participants based on sequence length or attention distribution to reduce communication costs and improve response quality.

	\item \textit{From a system perspective}, techniques such as bandwidth scheduling for resource allocation, differential privacy mechanisms, and Byzantine-robust aggregation for adversarial resilience can enhance FedAttn's efficiency, privacy, and robustness in large-scale collaborative LLM tasks.
\end{enumerate}

\section{Preliminaries}\label{Revisiting Self-Attention Mechanism}
\subsection{The Transformer Architecture}
\begin{figure*}[!t]
	\centering
	\includegraphics[width=7in,angle=0]{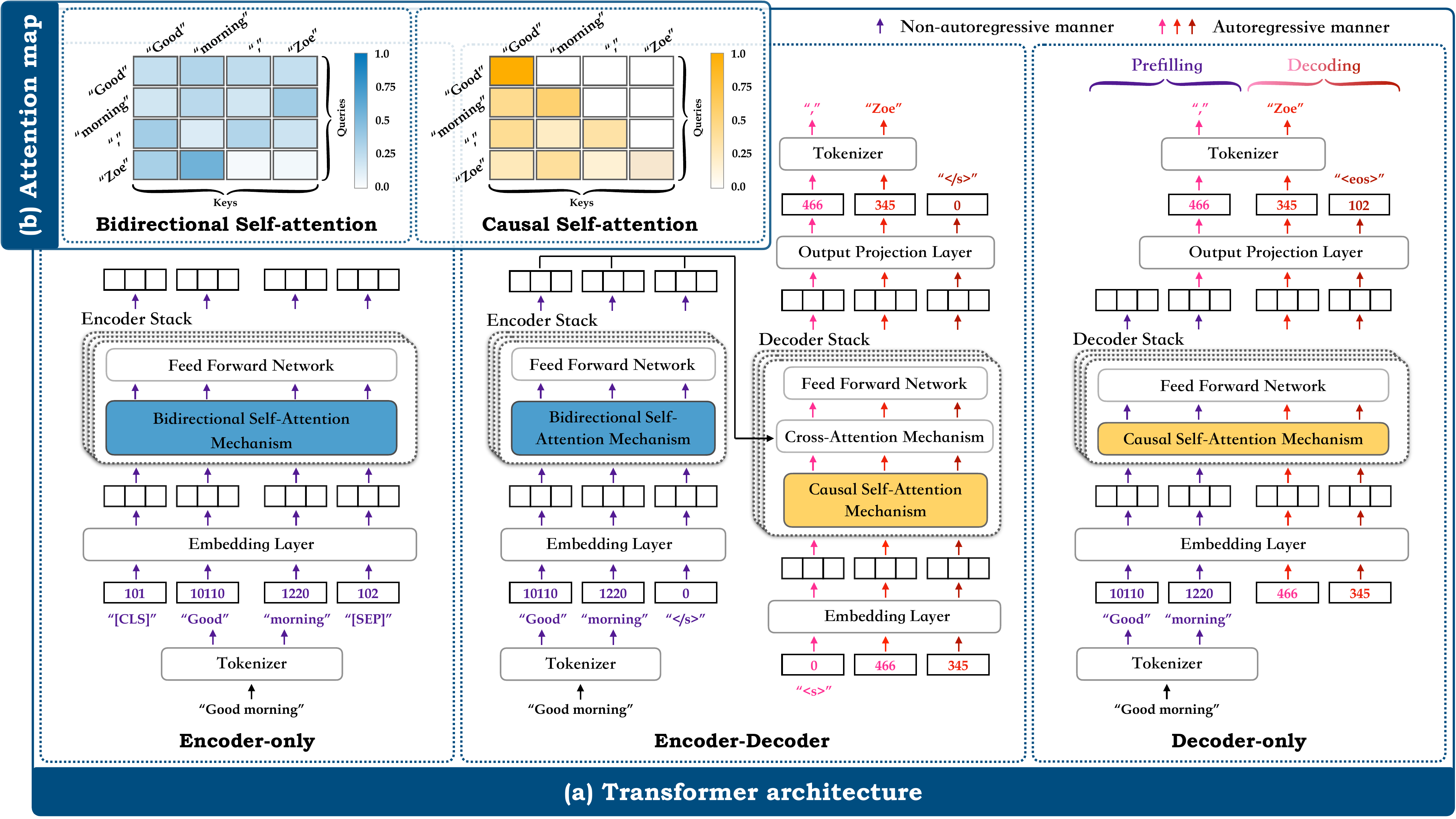}\vspace{5pt}
	\caption{Illustration of Transformer architecture and Attention map.}
	\label{fig_pre}
\end{figure*}
The Transformer architectures form the backbone of modern LLMs, categorized into three families: \textit{1) Encoder-only architecture} for contextual representation learning; \textit{2) Decoder-only architecture} for text generation, powering today's frontier LLMs like GPT series; \textit{3) Encoder-Decoder architecture} combining both text understanding and generating capabilities for tasks like translation. 

The Encoder-Decoder architecture defines the original Transformer through two key components, i.e., an encoder stack and a decoder stack. As illustrated in Fig.~\ref{fig_pre}, its inference process commences with a tokenizer that partitions the input sequence into discrete units called tokens, which are subsequently mapped to numerical identifiers (IDs) according to a vocabulary. An input embedding layer then transforms these token IDs into dense vector representations, which are combined with positional encodings to capture sequential information. The encoder stack then passes these token representations one encoder block after another, with each processing its hidden representations in a non-autoregressive manner through a bidirectional self-attention mechanism where tokens attend to each other followed by a position-wise feed-forward network (FFN). The decoder stack then executes a decoding process that is kickstarted by feeding a special beginning-of-sequence (BOS) token as its initial input, and followed by iterative decoding steps generating output tokens one by one in an autoregressive manner. In each decoding iteration, the decoder stack processes its input token with each decoder passing the hidden representation through a causal self-attention mechanism attending to previously generated output tokens, a cross-attention attending to the encoder's output, and an FFN transformation. The output projection layer transforms the final decoder output into a token probability distribution for predicting the next token ID, which is subsequently converted to text via the tokenizer and fed as input to the decoder stack in the next iteration.

Building upon the Encoder-Decoder architecture, the Encoder-only variant retains only the encoder components, directly feeding hidden representations from the encoder stack to an output projection layer. The Decoder-only variant retains only the decoder stack and operates with a Prefilling stage followed by a Decoding stage. In the Prefilling stage, the decoder stack processes all input tokens non-autoregressively using a causal self-attention mechanism, where each token can only attend to preceding tokens rather than the bidirectional attention in encoders. This difference arises from that to learn to predict the next token, decoders cannot access future tokens otherwise they would simply learn to copy rather than predict. In the Decoding stage, the decoder stack then generates output tokens autoregressively one at a time through causal self-attention to attend to both input tokens and previously generated tokens.

\subsection{Vanilla Self-Attention Mechanism}
Following `Attention Is All You Need'\cite{DBLP:conf/nips/VaswaniSPUJGKP17}, the vanilla self-attention mechanism in the Transformer is defined as\begin{align}\text{Attention}\left(\boldsymbol{Q},\boldsymbol{K},\boldsymbol{V}\right)=\text{Softmax}\left(\frac{\boldsymbol{Q}\boldsymbol{K}^\top}{\sqrt{d}}\right)\boldsymbol{V},\end{align}where $\boldsymbol{Q}$, $\boldsymbol{K}$, and $\boldsymbol{V}$ denote the Query, Key, and Value matrices, respectively, with each row corresponding to a contextualized feature vector of an input token projected into distinct representational subspaces. Specifically, $\boldsymbol{Q}$ encodes the attention patterns determining what information each token seeks from other positions, $\boldsymbol{K}$ describes what information each token can provide, and $\boldsymbol{V}$ contains the processed information that each token contributes to the attention output, respectively. This scaled dot-product attention operates in three steps within each Transformer block: \textit{1) {Query-Key-Value (QKV) Projection.}} Create $\boldsymbol{Q}$, $\boldsymbol{K}$, and $\boldsymbol{V}$ from input tokens, i.e.,\begin{align}\label{eq:QKV}
	\boldsymbol{Q}=\boldsymbol{X}\boldsymbol{W}_{Q}, \quad
	\boldsymbol{K}=\boldsymbol{X}\boldsymbol{W}_{K},\quad
	\boldsymbol{V}=\boldsymbol{X}\boldsymbol{W}_{V}, 
\end{align}where $\boldsymbol{X}\in\mathbb{R}^{L\times d}$ represents the hidden representations with $L$ being the sequence length and $d$ the hidden dimension, and $\boldsymbol{W}_{Q}$, $\boldsymbol{W}_{K}$, $\boldsymbol{W}_{V}\in\mathbb{R}^{d\times d}$ denote the learnable weight matrices for the linear projections with bias vectors omitted for brevity. \textit{2) {Query-Key Dot Product.}} Compute attention weights by applying a softmax function to the scaled dot-products between queries and keys, i.e.,\begin{align}\boldsymbol{A}=\text{Softmax}\left(\frac{\boldsymbol{Q}\boldsymbol{K}^\top}{\sqrt{d}}\right)\in\mathbb{R}^{L\times L},\end{align}termed the attention map as illustrated in Fig.~\ref{fig_pre}(b). Each row yields a probability distribution that reflects how relevant Keys are to each Query, answering ``Given what this token is seeking, how relevant is what each other token can provide?" \textit{3) {Value Aggregation.}} Obtain attention output by aggregating Values with the attention weights, producing token representations that capture long-range dependencies and contextual information across the complete sequence.

Building upon this bidirectional self-attention mechanism, causal self-attention is implemented by masking the attention weights with a causal mask, i.e., \begin{align}\tilde{\boldsymbol{A}}=\text{Softmax}\left(\frac{\boldsymbol{Q}\boldsymbol{K}^\top}{\sqrt{d}}+ \boldsymbol{M}\right),\end{align}where the causal mask $\boldsymbol{M}\in\mathbb{R}^{L\times L}$ is defined as\begin{equation}
\left(\boldsymbol{M}\right)_{i,j} = 
\begin{cases}
	0,&{\text{if}}\ i\leq j,\\ 
	{-\infty,}&{\text{otherwise.}} 
\end{cases}
\end{equation}

A complete Transformer block passes hidden representations through a self-attention mechanism followed by an FFN, with both sub-layers applying residual connections and layer normalization, as follows:\begin{equation}\boldsymbol{X}_\text{res} = \boldsymbol{X} + \text{SubLayer}\left(\boldsymbol{X}\right)\end{equation}with $\text{SubLayer}\left(\cdot\right)$ denoting either the self-attention mechanism or FFN, and $\boldsymbol{X}_\text{res} \in \mathbb{R}^{L\times d}$ representing the residual hidden representations, and\begin{equation}\left(\text{LN}(\boldsymbol{X})\right)_{l,i} = \left(\boldsymbol{\gamma}\right)_i \frac{\left(\boldsymbol{X}\right)_{l,i} - \mu_{l}}{\sqrt{\sigma^2_{l} + \epsilon}} + \left(\boldsymbol{\beta}\right)_i,\end{equation}where $\boldsymbol{\gamma}, \boldsymbol{\beta} \in \mathbb{R}^d$ are learnable scale and shift parameters, and $\mu_{l}, \sigma^2_{l}$ denote the mean and variance computed over each row vector $\left(\boldsymbol{X}\right)_{l,:}$. Taking the Pre-Normalization (Pre-LN) architecture as an example, which applies layer normalization before each sub-layer within the residual branch, the final output of a Transformer block is computed as\begin{align}
	\boldsymbol{X}_{\text{Attn}\_\text{Res}} = &\boldsymbol{X}_\text{in} +\text{Attention}\Big(\text{LN}\left(\boldsymbol{X}_\text{in}\right)\boldsymbol{W}_{Q},\nonumber\\&\text{LN}\left(\boldsymbol{X}_\text{in}\right)\boldsymbol{W}_{K},\text{LN}\left(\boldsymbol{X}_\text{in}\right)\boldsymbol{W}_{V}\Big), \label{eq:layer_attn}
	\end{align}followed by\begin{align}
	\boldsymbol{X}_{\text{out}} = \boldsymbol{X}_{\text{Attn}\_\text{Res}} + \text{FFN}\Big(\text{LN}\left(\boldsymbol{X}_{\text{Attn}\_\text{Res}}\right)\Big)\label{eq:layer_ffn},
\end{align}where $\text{Attention}(\cdot)$ uses a slight abuse of notation to denote either bidirectional or masked self-attention mechanisms, $\boldsymbol{X}_{\text{in}}$ and $\boldsymbol{X}_{\text{out}}$ denote the input and output hidden representations of the Transformer block, $\boldsymbol{X}_{\text{Attn}\_\text{Res}}$ denote the attention output with residual connection, and\begin{align}\text{FFN}(\boldsymbol{X})=\sigma\Big(\boldsymbol{X}\boldsymbol{W}_\text{ffn1}\Big)\boldsymbol{W}_\text{ffn2},\end{align}represents the position-wise FFN with learnable weight matrices $\boldsymbol{W}_\text{ffn1}\in\mathbb{R}^{d\times d_\text{ff}}$, $\boldsymbol{W}_\text{ffn2}\in\mathbb{R}^{d_\text{ff}\times d}$, and a non-linear activation function $\sigma(\cdot)$ such as ReLU.

\subsection{Computational complexity}
Self-attention presents a dominant computational bottleneck in long-context LLM inference, primarily stemming from the non-autoregressive components of Transformer architectures, i.e., the encoder and decoder’s Prefilling stage. Specifically, the floating-point operations (FLOPs) and memory requirements mainly arise from the Query-Key dot product operation, yielding quadratic computational and memory complexity that scales as $\mathcal{O}(L d^2 + L^2 d)$ and $\mathcal{O}(L^2 + L d)$, respectively. 

To avoid redundant computation, Autoregressive decoding employs KV caching to retain all preceding KV pairs throughout the inference process, reducing computational complexity to $\mathcal{O}(L)$ per decoding step but introducing a significant memory bottleneck as KV caches accumulate linearly with each generated token.

\section{Federated Attention Paradigm}\label{Federated Attention Paradigm}
In this section, we present the federated attention paradigm, which is tailored for the non-autoregressive components of Transformer architectures. FedAttn enables multiple participants to collaborate to generate an LLM response without sharing their private prompts, by executing self-attention mechanism on their local token representations and exchanging KV pairs every few transformer blocks. 
\subsection{Problem Formulation and Notation}

\begin{figure}[!t]
	\centering
	\includegraphics[width=3.45in,angle=0]{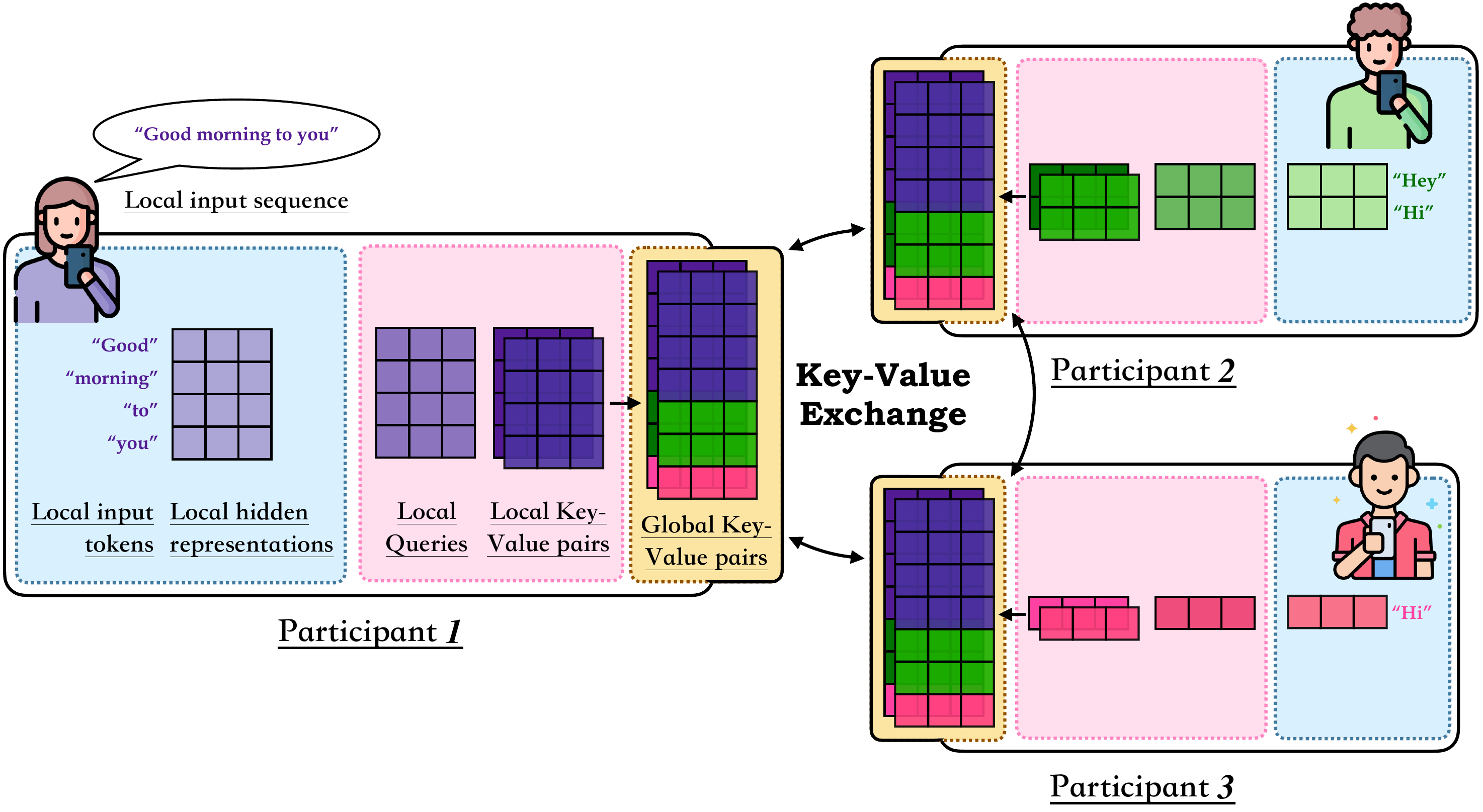}
	\caption{Framework Overview of FedAttn. We exemplify FedAttn through three participants, each maintaining private input tokens while collaboratively computing attention through periodic KV exchange.}
	\label{fig0}
\end{figure}

As illustrated in Fig.~\ref{fig0}, we consider $N$ participants collaborating to perform a Transformer-based LLM inference task, where each participant contributes their respective private inputs with one participant acting as the task publisher by issuing the query and receiving the final output, and the other $N-1$ participants providing local prompts such as records and domain-specific documents relevant to answering this query. Each participant begins with passing its private prompt through tokenizer and embedding layer to obtain input token embeddings which are then fed to the Transformer stack.

Following FL conventions, we refer to the aggregations of all participants' local input sequences, tokens, and embeddings as the global input sequence, global input tokens, and global input embeddings, respectively. Let $L$ denote the length of global input sequence, $\mathcal{L}=(1,2,...,L)$ denote the index set of global input tokens, and \begin{align}\boldsymbol{X}^\text{emb}=\begin{bmatrix}\boldsymbol{X}^\text{emb}_{1,:}\\\boldsymbol{X}^\text{emb}_{2,:}\\\vdots\\\boldsymbol{X}^\text{emb}_{L,:}\end{bmatrix}\in\mathbb{R}^{L\times d}\end{align}denote the global input embeddings, respectively. For the $n$-th participant, let $L_n$ denote the length of local input sequence, and $\mathcal{L}_n=\left(i_n^{j}\right)_{j=1}^{L_n}\subset\mathcal{L}$ denote the index set of local input tokens with the ordering $i_n^{1} < i_n^{2} < \dots < i_n^{L_n}$. Notably, $\left\{\mathcal{L}_n\right\}_{n=1}^N$ constitute a disjoint partition of $\mathcal{L}$, i.e., $L=\sum_{n=1}^{N}L_n$, and $\mathcal{L}_n \bigcap \mathcal{L}_{n'} = \emptyset$ for all $n \ne n'$. Define an $L\times L_n$ binary indicator matrix\begin{align}\boldsymbol{\Pi}_n=\begin{bmatrix} \boldsymbol{e}_{i_n^{1}}&\boldsymbol{e}_{i_n^{2}}&\cdots&\boldsymbol{e}_{i_n^{L_n}}\end{bmatrix},\end{align}where the $j$-th column $\boldsymbol{e}_{i_n^{j}}$ is the $i_n^{j}$-th standard basis vector of $\mathbb{R}^L$. The local input embeddings \begin{align}\boldsymbol{X}_n^{\text{emb}}=\begin{bmatrix} \boldsymbol{X}^{\text{emb}}_{i_n^{1},:}\\\boldsymbol{X}^{\text{emb}}_{i_n^{2},:}\\\vdots\\\boldsymbol{X}^{\text{emb}}_{i_n^{L_n},:}\end{bmatrix}\in\mathbb{R}^{L_n\times d}\end{align}can be written as \begin{align}\boldsymbol{X}_n^{\text{emb}}=\boldsymbol{\Pi}_n^\top \boldsymbol{X}^{\text{emb}},\end{align}where globally \begin{align}\boldsymbol{X}^{\text{emb}}=\sum_{n=1}^{N}\boldsymbol{\Pi}_n \boldsymbol{X}_n^{\text{emb}}.\end{align}
\subsection{Algorithmic Procedure}
\begin{figure*}[!t]
	\centering
	\includegraphics[width=7.3in,angle=0]{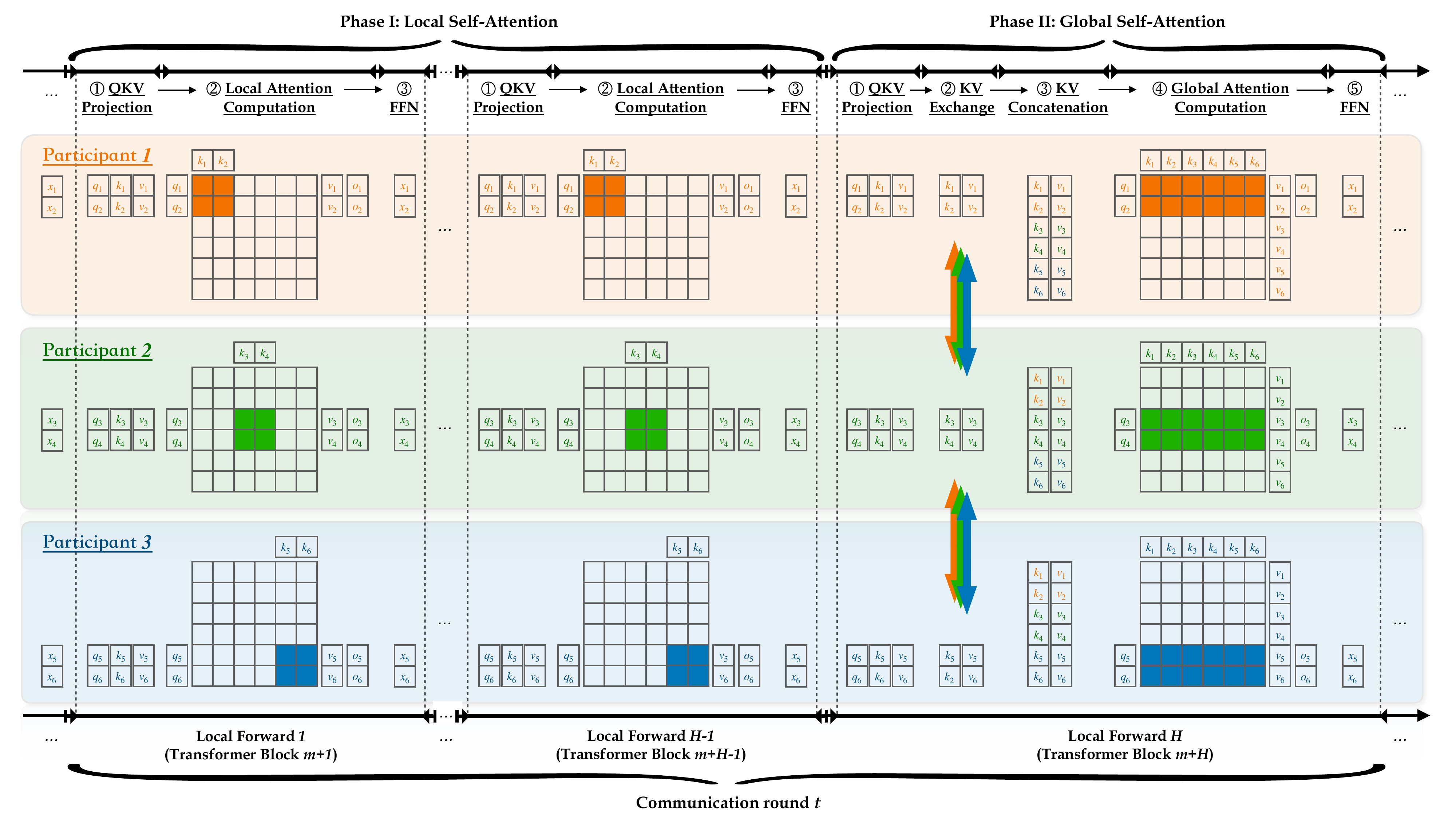}
	\caption{Algorithmic Procedure of FedAttn. We illustrate FedAttn through a representative example of one communication round involving three participants, each with two local input tokens, executing Transformer-based LLM inference. For expository clarity, the notations for hidden representations, Queries, Keys, and Values are streamlined as vectors indexed by token IDs as subscripts.}
	\label{fig1}
\end{figure*}
Following FL conventions, we refer to a forward pass of local hidden representaitons through a Transformer block as a local forward, with multiple successive local forwards comprising one communication round. In this paper, we consider a synchronous setting where all FedAttn participants exchange their local KV matrices within each communication round. Denote the index sets of local forwards and communication rounds by $\mathcal{H}=\{1,2,\ldots,H\}$ and $\mathcal{T}=\{0,1,\ldots,T-1\}$, respectively. We consider $M = HT$ Transformer blocks indexed by $\mathcal{M}=\{1,2,\ldots,M\}$, where $m = Ht + h$. Denote the learnable parameters of the $m$-th block by $\boldsymbol{\Theta}^m = \left\{\boldsymbol{W}_{Q}^m, \boldsymbol{W}_{K}^m, \boldsymbol{W}_{V}^m, \boldsymbol{W}_\text{ffn1}^m, \boldsymbol{W}_\text{ffn2}^m\right\}$ with bias and normalization parameters omitted for brevity.

\begin{algorithm}[!t]
	\SetKwBlock{DoParallel}{in parallel}{end}
	\SetKw{kwInit}{Init:}
	\caption{FedAttn}\label{alg:FedAttn}
	\KwIn{Local input embeddings $\{\boldsymbol{X}_n^{\text{emb}}\}_{n=1}^N$}
	\KwOut{Global hidden representations $\boldsymbol{X}^{T}$}
	Initialize local hidden representations $\left\{\boldsymbol{x}_n^{1,0}\right\}_{n=1}^N$ according to (\ref{eq:init_kv})\\
	\For{\rm{each communication round} $t\in \{0,1,...,T-1\}$}{
		\For(\textbf{in parallel}){\rm{each participant} $n\in \{1,2,...,N\}$}{
			\For{\rm{each local forward} $h\in \{1,2,...,H-1\}$}{
				Initialize local Queries, Keys, and Values $\boldsymbol{q}_n^{h,t},\boldsymbol{k}_n^{h,t},\boldsymbol{v}_n^{h,t}$ via (\ref{eq:update_local_qkv});\\
				Compute local attention output $\boldsymbol{o}_n^{h,t}$ according to (\ref{eq:update_local_o});\\
				Update local hidden representations $\boldsymbol{x}_n^{h,t}$ according to (\ref{eq:update_local_x})\\}
			\For{\rm{local forward} $h=H$}{
				Initialize local Queries, Keys, and Values $\boldsymbol{q}_n^{h,t},\boldsymbol{k}_n^{h,t},\boldsymbol{v}_n^{h,t}$ via (\ref{eq:update_local_qkv})\\}						
		}
		All participants exchange local KV matrices $\begin{bmatrix}\boldsymbol{k}_n^{H,t}\;\boldsymbol{v}_n^{H,t}\end{bmatrix}$ and obtain the global KV matrix $\begin{bmatrix}\boldsymbol{K}^{t}\;\boldsymbol{V}^{t}\end{bmatrix}$ according to (\ref{eq:update_global_kv_H})\\
		\For(\textbf{in parallel}){\rm{each participant} $n\in \{1,2,...,N\}$}{
			\For{\rm{local forward} $h=H$}{
				Compute local attention output $\boldsymbol{o}_n^{h,t}$ according to (\ref{eq:update_local_o_H});\\
				Update local hidden representations $\boldsymbol{x}_n^{1,t+1}$ according to (\ref{eq:update_local_x})\\}
		}
	}
\end{algorithm}

As depicted in Algorithm \ref{alg:FedAttn}, each participant in FedAttn begins by initializing local hidden representations with its respective local input embeddings, i.e.,\begin{align}\label{eq:init_kv}\boldsymbol{x}_n^{1,0}=\boldsymbol{X}_n^\text{emb},\end{align}followed by passing them through the Transformer block stack. The algorithmic procedure of FedAttn for each communication round is detailed as follows.

\textit{\textbf{Phase I: Local Self-Attention.} }At the beginning of each communication round, each participant passes its local hidden representations through $H-1$ Transformer blocks, with each block beginning with Query-Key-Value projection, i.e.,\begin{align}\label{eq:update_local_qkv}\begin{bmatrix}\boldsymbol{q}_n^{h,t}\;\boldsymbol{k}_n^{h,t}\; \boldsymbol{v}_n^{h,t}\end{bmatrix}=\text{LN}\Big(\boldsymbol{x}_n^{h,t}\Big)\begin{bmatrix}\boldsymbol{W}_{Q}^m\;\boldsymbol{W}_{K}^m\; \boldsymbol{W}_{V}^m\end{bmatrix},\end{align}where $\boldsymbol{x}_n^{h,t}$ denotes the local hidden representations of the $n$-th participant in the $h$-th local forward of the $t$-th communication round, and $\boldsymbol{q}_n^{h,t}$, $\boldsymbol{k}_n^{h,t}$, and $\boldsymbol{v}_n^{h,t}$ denote the local Query, Key, and Value matrix, respectively, followed by computing the local attention output, i.e.,\begin{align}\label{eq:update_local_o}\boldsymbol{o}_n^{h,t}=\text{Attention}\left(\boldsymbol{q}_n^{h,t},\boldsymbol{k}_n^{h,t},\boldsymbol{v}_n^{h,t}\right).\end{align}Subsequent to the self-attention mechanism, the local hidden representations are transformed through residual connections with the local attention output followed by an FFN, according to the update rule, i.e.,\begin{align}\label{eq:update_local_x}\boldsymbol{x}_n^{h+1,t}=\boldsymbol{x}_n^{h,t}+\boldsymbol{o}_n^{h,t}+\text{FFN}\left(\text{LN}\left(\boldsymbol{x}_n^{h,t}+\boldsymbol{o}_n^{h,t}\right)\right).\end{align}

\textit{\textbf{Phase II: Global Self-Attention.} }Upon completing the $H-1$ local forwards, each participant passes its respective hidden representations $\boldsymbol{x}_n^{H,t}$ through the subsequent $H$-th Transformer block, beginning by projecting $\boldsymbol{x}_n^{H,t}$ into Query, Key, and Value matrices $\begin{bmatrix}\boldsymbol{q}_n^{H,t}\;\boldsymbol{k}_n^{H,t}\; \boldsymbol{v}_n^{H,t}\end{bmatrix}$. Subsequently, all participants exchange local KV matrices and aggregate them to obtain the global KV matrix, i.e.,\begin{align}\label{eq:update_global_kv_H}\begin{bmatrix}\boldsymbol{K}^{t}\;\boldsymbol{V}^{t}\end{bmatrix}=\sum_{n=1}^{N}\boldsymbol{\Pi}_n \begin{bmatrix}\boldsymbol{k}_n^{H,t}\;\boldsymbol{v}_n^{H,t}\end{bmatrix}.\end{align}Each participant then computes local attention output where local Query matrix attends to global KV matrix, i.e.,\begin{align}\label{eq:update_local_o_H}\boldsymbol{o}_n^{H,t}=\text{Attention}\left(\boldsymbol{q}_n^{H,t},\boldsymbol{K}^{t},\boldsymbol{V}^{t}\right),\end{align} followed by applying FFN transformation to update the local hidden representations $\boldsymbol{x}_n^{1,t+1}$.

\subsection{Output Generation}

Upon completing the $T$ communication rounds in FedAttn, all participants have collaboratively completed the non-autoregressive components of this LLM inference task, with each encoding its local input sequence with global contextual information integrated from other participants, and producing globally augmented local token representations including hidden representations and KVs. Subsequent to this, the task publisher generates the final outputs of this LLM inference task with specific operations depending on task objectives and Transformer architectures.

For encoder-only architectures, the task publisher feeds its local hidden representations $\boldsymbol{x}_n^{1,T}$ into the output projection layer to produce predictions such as classification logits and semantic similarity scores in downstream understanding tasks. For encoder-decoder and decoder-only architectures, FedAttn maintains KV caches at each Transformer block for reuse in the subsequent Decoding stage, where blocks executing local self-attention cache local KV matrices and those performing global self-attention cache the global KV matrix. The task publisher begins the autoregressive decoding from the BOS token with encoder-decoder LLMs, or from the final token of the global input sequence with decoder-only LLMs, generating the output response in text generation tasks.

\section{Federated Duality: From FL to FedAttn}\label{Federated Duality: From FL to FedAttn}
To elucidate the key insights of FedAttn, we examine its duality with FL in this section, focusing on how federated paradigm is integrated into self-attention mechanisms.

We begin by examining FedAttn's local self-attention computation that restricts Queries attending only to local KV pairs, as formalized by rewriting (\ref{eq:update_local_o}) as\begin{align}\label{reeq:update_local_o}
	\boldsymbol{o}_n^{h,t}\equiv \text{Attn}\left(\boldsymbol{q}_n^{h,t}\middle|\boldsymbol{k}_n^{h,t}, \boldsymbol{v}_n^{h,t}\right).
\end{align}Define an auxiliary notation $\hat{\boldsymbol{o}}_n^{h,t}$ representing the attention output of the global self-attention counterpart where Queries attend to global KV pairs, expressed by\begin{align}
	\hat{\boldsymbol{o}}_n^{h,t} = \text{Attn}\left(\boldsymbol{q}_n^{h,t}\middle|{\boldsymbol{K}}^{h,t}, {\boldsymbol{V}}^{h,t}\right),
\end{align}with\begin{align}\begin{bmatrix}{\boldsymbol{K}}^{h,t}\; {\boldsymbol{V}}^{h,t}\end{bmatrix}=\sum_{n=1}^{N}\boldsymbol{\Pi}_n \begin{bmatrix}\boldsymbol{k}_n^{h,t}\; \boldsymbol{v}_n^{h,t}\end{bmatrix}\end{align}representing the global KV matrix in the $h$-th local forward of the $t$-th communication round.\begin{observation}[\textnormal{Local versus Global Self-Attention}]\label{O1}The deviation introduced by the local self-attention computation can be characterized by\begin{align}
		\left\Vert\boldsymbol{o}_n^{h,t}-\hat{\boldsymbol{o}}_n^{h,t}\right\Vert_F = \Big\Vert& \text{Attn}\left(\boldsymbol{q}_n^{h,t}\middle|\boldsymbol{k}_n^{h,t}, \boldsymbol{v}_n^{h,t}\right)\nonumber\\&-\text{Attn}\left(\boldsymbol{q}_n^{h,t}\middle|{\boldsymbol{K}}^{h,t}, {\boldsymbol{V}}^{h,t}\right)\Big\Vert_F,
	\end{align}where $\Vert\cdot\Vert_F$ denotes the Frobenius norm. This deviation diminishes as attention map exhibits increasing locality, where Queries demonstrate higher relevance to local Keys $\boldsymbol{k}_n^{h,t}$, and thereby attend more to local Values $\boldsymbol{v}_n^{h,t}$ than to those from other participants.
\end{observation} 

We now proceed to compare FedAttn with FL in Tab. \ref{tab:FL2FedAttn} across three key components of \textit{1) {private data}}, \textit{2) {local computation}}, and \textit{3) {global aggregation}}. The notations of the two federated paradigms are defined as follows. Within FL, let $\mathcal{D}_n$ denote the local training dataset of the $n$-th participant with $D_n=\vert\mathcal{D}_n\vert$ data points, each comprising a feature vector $\boldsymbol{x}_{n,i} \in\mathbb{R}^d$ and its corresponding label $y_{n,i}\in\mathbb{R}$. Let $\mathcal{D} = \bigcup_{n=1}^{N} \mathcal{D}_n$ denote the global training dataset with a total of $D = \sum_{n=1}^{N} D_n$ data points. Let $\boldsymbol{w}_n^{h,t}$ denote the local model parameters of the $n$-th participant in the $h$-th local epoch of the $t$-th communication round, which are updated via gradient descent optimization algorithm as\begin{align}\boldsymbol{w}_n^{h+1,t}=\boldsymbol{w}_n^{h,t}-\eta\,\boldsymbol{g}_n^{h,t},\end{align}where $\eta$ is the step size, and $\boldsymbol{g}_n^{h,t}$ denotes the local gradient, given by\begin{align}\boldsymbol{g}_n^{h,t}=\nabla f\left(\boldsymbol{w}_n^{h,t}\middle| \mathcal{D}_n\right)\end{align}with $f\left(\boldsymbol{w}\middle| \mathcal{D}_n\right)$ representing the local loss function. The accumulated local gradients are aggregated to obtain a global gradient as given by\begin{align}\boldsymbol{G}^t=\sum_{n=1}^N \alpha_n\left(\sum_{h=1}^H\boldsymbol{g}_n^{h,t}\right),\end{align}with $\alpha_n$ denoting the averaging weight for the $n$-th participant. The global model parameters are then updated by\begin{align}\boldsymbol{W}^{t+1}=\boldsymbol{W}^{t}-\eta\boldsymbol{G}^t.\end{align}For ease of exposition, we rewrite FedAttn's local update rule for hidden representations from (\ref{eq:update_local_x}) as\begin{align}\boldsymbol{x}_n^{h+1,t}\equiv\boldsymbol{x}_n^{h,t}+\boldsymbol{o}_n^{h,t}+\mathcal{F}^m\left(\boldsymbol{x}_n^{h,t}+\boldsymbol{o}_n^{h,t}\right),\end{align}where $\mathcal{F}^m\left(\cdot\right)$ represents the position-wise FFN operator with layer normalization of the $m$-th Transformer block, which is applied independently to each token representation. We rewrite the Query-Key-Value projection in (\ref{eq:update_local_qkv}) as\begin{align}\begin{bmatrix}\boldsymbol{q}_n^{h,t}\;\boldsymbol{k}_n^{h,t}\; \boldsymbol{v}_n^{h,t}\end{bmatrix}=\mathcal{P}^m_\text{QKV}(\boldsymbol{x}_n^{h,t}),\end{align}where \begin{align}\mathcal{P}^m_\text{QKV}(\cdot)=\begin{bmatrix}\mathcal{P}^m_\text{Q}(\cdot)\;\mathcal{P}^m_\text{K}(\cdot)\; \mathcal{P}^m_\text{V}(\cdot)\end{bmatrix}\end{align}defines the position-wise Query-Key-Value projection operators with layer normalization for the $m$-th Transformer block.

\begin{table*}[!t]
	\caption{A Federated Duality between FL and FedAttn\label{tab:FL2FedAttn}}
	\centering
	\renewcommand{\arraystretch}{1.75}
	\begin{tabular}{!{\vrule width 0pt}ccccc!{\vrule width 0pt}}
		\noalign{\hrule height 1pt}
		~&\multicolumn{2}{c}{{FL}}&\multicolumn{2}{c}{{FedAttn}}\\\cline{2-5}
		~&Description&Notation&Description&Notation\\\noalign{\hrule height 1pt}
		Private Data&Training datasets& $\left\{\mathcal{D}_n \right\}_{n=1}^N$&Input tokens&$\left\{\mathcal{L}_n\right\}_{n=1}^N$\\	
		\noalign{\hrule height 0.5pt}
		\multirow{8}*{\!\!\makecell[c]{Local \\Computation}\!\!\!\!}&Model parameters&$\boldsymbol{w}_n^{h,t}$&Hidden representations&$\boldsymbol{x}_n^{h,t}$\\\cline{2-5}
		~&Loss function&$f\left(\boldsymbol{w}_n^{h,t}\middle| \mathcal{D}_n\right)$&Self-attention mechanism&$\text{Attn}\left(\boldsymbol{q}_n^{h,t}\middle| \boldsymbol{k}_n^{h,t}, \boldsymbol{v}_n^{h,t}\right)$\\\cline{2-5}
		~&\multicolumn{2}{l}{\multirow{3}*{\makecell[c]{\vspace{-0.02cm}\\\parbox[t]{6cm}{\raggedright {\raggedright {\tiny\ding{110}\;} Local Gradient Descent (Backward passes, \\\quad$h = 1, \ldots, H$):}}}}}\vspace{-0.12cm}&\multicolumn{2}{l}{\parbox[t]{8cm}{\raggedright {{\tiny\ding{110}\;} Local Self-Attention (Forward passes, $h = 1, \ldots, H-1$):}}}\\	
		
		~&~&~&\multirow{2}*{\;\;\;\parbox[t]{3.1cm}{\raggedright {\bf{\tiny{\circled{1}}}} Query-Key-Value projection}}&\multirow{2}*{$\begin{bmatrix}\boldsymbol{q}_n^{h,t}\;\boldsymbol{k}_n^{h,t}\; \boldsymbol{v}_n^{h,t}\end{bmatrix}=\mathcal{P}^m_\text{QKV}\left(\boldsymbol{x}_n^{h,t}\right)$}\\~&~\vspace{-0.3cm}&~&~&~\\
		
		~&\multirow{2}*{\!\!\!\parbox[t]{2.1cm}{\raggedright {\bf{\tiny{\circled{1}}}} Local {gradient\!\!\!\!} computation\!\!\!\!\!\!}}&\multirow{2}*{$\boldsymbol{g}_n^{h,t}=\nabla f\left(\boldsymbol{w}_n^{h,t}\middle|\mathcal{D}_n\right)$}&\multirow{2}*{\;\;\;\parbox[t]{3.1cm}{\raggedright {{\bf{\tiny{\circled{2}}}}} Local self-attention computation}}&\multirow{2}*{$\boldsymbol{o}_n^{h,t}=\text{Attn}\left( \boldsymbol{q}_n^{h,t}\middle|\boldsymbol{k}_n^{h,t}, \boldsymbol{v}_n^{h,t}\right)$}\\~&~\vspace{-0.3cm}&~&~&~\\	
		
		~&\multirow{2}*{\!\!\!\parbox[t]{2.1cm}{\raggedright {\bf{\tiny{\circled{2}}}} Local model optimization}}&\multirow{2}*{$\boldsymbol{w}_n^{h+1,t}=\boldsymbol{w}_n^{h,t}-\eta\,\boldsymbol{g}_n^{h,t}$}&\multirow{2}*{\;\;\;\parbox[t]{3.1cm}{\raggedright {\bf{\tiny{\circled{3}}}} Local hidden representation refinement\!\!\!}}&\multirow{2}*{\makecell[l]{$\boldsymbol{x}_n^{h+1,t}=\boldsymbol{x}_n^{h,t}+\boldsymbol{o}_n^{h,t}+\mathcal{F}^m\left(\boldsymbol{x}_n^{h,t}+\boldsymbol{o}_n^{h,t}\right)$}}\\~&~\vspace{-0.1cm}&~&~&~\\
		
		\noalign{\hrule height 0.5pt}
		
		\multirow{7}*{\!\!\!\makecell[c]{Global \\Aggregation}\!\!\!\!}&~&~\vspace{-0.12cm}&\multicolumn{2}{l}{\parbox[t]{9.2cm}{\raggedright {{\tiny\ding{110}\;}  Global Self-Attention (Forward pass, $h=H$):}}}\\			
		
		~&\multicolumn{2}{l}{\multirow{2}*{\vspace{0.05cm}\makecell[c]{\\\parbox[t]{6cm}{\raggedright {\raggedright {\tiny\ding{110}\;} Global Gradient Descent:}}}}}&~\vspace{0.08cm}\multirow{2}*{\;\;\parbox[t]{3.1cm}{\raggedright {\bf{\tiny{\circled{1}}}} Query-Key-Value projection}}&\multirow{2}*{$\begin{bmatrix}\boldsymbol{q}_n^{H,t}\;\boldsymbol{k}_n^{H,t}\; \boldsymbol{v}_n^{H,t}\end{bmatrix}=\mathcal{P}^m_\text{QKV}\left(\boldsymbol{x}_n^{H,t}\right)$}\\~&\multicolumn{2}{l}{\multirow{2}*{\;\parbox[t]{5.2cm}{\raggedright {\bf{\tiny{\circled{1}}}} Gradient exchange}}}&\multirow{2}*{\;\;\;\parbox[t]{3.1cm}{\raggedright {\bf{\tiny{\circled{2}}}} KV exchange}}&~\\~&~\vspace{-0.5cm}&~&~&~\\
		
		~&\multirow{3}*{\vspace{-0.5em}\;\parbox[t]{2.9cm}{\raggedright {\bf{\tiny{\circled{2}}}} Gradient aggregation \\\vspace{0.2em}(Global gradient computation)}}&\multirow{3}*{$\boldsymbol{G}^t=\sum\limits_{n=1}^N \alpha_n\left(\sum\limits_{h=1}^H\boldsymbol{g}_n^{h,t}\right)$}&\multirow{2}*{\;\;\;\parbox[t]{3.1cm}{\raggedright {\bf{\tiny{\circled{3}}}} KV Aggregation}}&\multirow{2}*{$\begin{bmatrix}\boldsymbol{K}^{t}\; \boldsymbol{V}^{t}\end{bmatrix}=\sum_{n=1}^{N}\boldsymbol{\Pi}_n \begin{bmatrix}\boldsymbol{k}_n^{H,t}\; \boldsymbol{v}_n^{H,t}\end{bmatrix}$}\\~&~\vspace{-0.3cm}&~&~&~\\
		
		~&~&~&\multirow{2}*{\;\;\;\parbox[t]{3.1cm}{\raggedright {\bf{\tiny{\circled{4}}}} Global {self-attention\!\!\!} computation\!\!\!}}&\multirow{2}*{$\boldsymbol{o}_n^{H,t}=\text{Attn}\left( \boldsymbol{q}_n^{H,t}\middle|\boldsymbol{K}^{t}, \boldsymbol{V}^{t}\right)$}\\~&~\vspace{-0.3cm}&~&~&~\\	
		~&\multirow{2}*{\;\parbox[t]{2.9cm}{\raggedright {\bf{\tiny{\circled{3}}}} Global model optimization}}&\multirow{2}*{$\boldsymbol{W}^{t+1}=\boldsymbol{W}^{t}-\eta\boldsymbol{G}^t$}&\multirow{2}*{\;\;\;\parbox[t]{3.1cm}{\raggedright {\bf{\tiny{\circled{5}}}} Global hidden representation refinement\!\!\!\!}}&\multirow{2}*{\makecell[l]{$\boldsymbol{x}_n^{0,t+1}=\boldsymbol{x}_n^{H,t}+\boldsymbol{o}_n^{H,t}+\mathcal{F}^m\left(\boldsymbol{x}_n^{H,t}+\boldsymbol{o}_n^{H,t}\right)$}}\\~&~\vspace{-0.1cm}&~&~&~\\			
		\noalign{\hrule height 1pt}	
	\end{tabular}
\end{table*}
As illustrated in Tab. \ref{tab:FL2FedAttn}, the duality between FedAttn and FL is detailed below.\begin{observation}[\textnormal{Local update rule}]FedAttn mirrors the FL procedure in the local updates, where FedAttn's forward passes through successive Transformer blocks refine hidden representations via local self-attention computations, forming a duality with FL's backward passes that iteratively optimize model parameters via local gradient descent. We now elaborate on this duality within each local update as follows:\begin{enumerate}[label=\bf\textit{\arabic*)}]
		\item In FL, each participant computes the local gradient $\boldsymbol{g}_n^{h,t}$ with respect to its local model parameters $\boldsymbol{w}_n^{h,t}$ on the local dataset $\mathcal{D}_n$, then updates the local model parameters through one step of gradient descent optimization with this local gradient. 
		\item Each FedAttn participant initializes local Queries, Keys and Values from its local hidden representations $\boldsymbol{x}_n^{h,t}$, computes local attention output $\boldsymbol{o}_n^{h,t}$ with respect to its local Queries $\boldsymbol{q}_n^{h,t}$ attending to its local KV pairs $\begin{bmatrix}\boldsymbol{k}_n^{h,t}\; \boldsymbol{v}_n^{h,t}\end{bmatrix}$, then updates its hidden representations through residual connections with this attention output followed by an FFN transformation.\end{enumerate} 

To reduce computation complexity, FL typically employs stochastic or mini-batch gradient descent, computing stochastic gradients over mini-batches $\tilde{\mathcal{D}}_n$ sampled from local datasets as given by\begin{align}\tilde{\boldsymbol{g}}_n^{h,t}=\nabla f\left(\tilde{\boldsymbol{w}}_n^{h,t}\middle| \tilde{\mathcal{D}}_n\right).\end{align}Correspondingly, FedAttn finds computational efficiency via \textbf{Sparse Self-Attention Mechanism}\cite{DBLP:conf/aaai/TangZWLXZ22} over sampled local Query-Key-Value pairs $\begin{bmatrix}\tilde{\boldsymbol{q}}_n^{h,t}\; \tilde{\boldsymbol{k}}_n^{h,t}\; \tilde{\boldsymbol{v}}_n^{h,t}\end{bmatrix}$ with sampling indices $\mathcal{I}_n \subseteq \{1, \dots, L_n\}$ such that $\tilde{L}_n=\vert\mathcal{I}_n\vert<L_n$, reducing the local computational complexity to $\mathcal{O}\left(\tilde{L}_n d^2 + \left(\tilde{L}_n\right)^2 d\right)$ and yielding the sparse local attention output\begin{align}\tilde{\boldsymbol{o}}_n^{h,t}\equiv \text{Attn}\left(\tilde{\boldsymbol{q}}_n^{h,t}\middle|\tilde{\boldsymbol{k}}_n^{h,t}, \tilde{\boldsymbol{v}}_n^{h,t}\right).\end{align}Theoretically, sparse attention can be further optimized to reduce the approximation error $\left\Vert\tilde{\boldsymbol{o}}_n^{h,t}-\hat{\boldsymbol{o}}_n^{h,t}\right\Vert_F$, by strategically selecting critical tokens through analyzing historical attention maps or using heuristics such as temporal recency for streaming contexts and attention sinks for initial tokens\cite{DBLP:conf/nips/ZaheerGDAAOPRWY20,DBLP:conf/iclr/XiaoTCHL24,DBLP:journals/tii/WangWJY25,DBLP:conf/naacl/HanWPX0JW24}.
	
\end{observation}
\begin{observation}[\textnormal{Data distribution}]Whether FedAttn refines hidden representations or FL optimizes model parameters, both federated paradigms achieve optimal performance when local updates align with their global counterparts, fundamentally determined by data distributions across participants as follows:\begin{enumerate}[label=\bf\textit{\arabic*)}]
		\item FL's local gradient $\nabla f(\boldsymbol{w}_n^{h,t}| \mathcal{D}_n)$ defines the steepest descent direction at model parameters $\boldsymbol{w}_n^{h,t}$ conditioned on local dataset $\mathcal{D}_n$ that determines what information the model extracts from local training data, whereas the global gradient learns from the complete training dataset $\mathcal{D}$. FL approaches centralized machine learning when local gradients align with the global gradient, i.e., $\nabla f\left(\boldsymbol{w}\middle | \mathcal{D}_n\right) =\nabla f\left(\boldsymbol{w}\middle | \mathcal{D}\right)$ under the {independent and identically distributed (IID)} data distribution. 
		\item FedAttn manifests as \textbf{Attention Distribution}, i.e., attention weights encode token relevance within and across participants. The local attention output represents a combination of contextual information that tokens retrieve from other positions determined by queries $\boldsymbol{q}_n^{h,t}$ attending to KV pairs $\begin{bmatrix}\boldsymbol{k}_n^{h,t}\; \boldsymbol{v}_n^{h,t}\end{bmatrix}$, where $\boldsymbol{q}_n^{h,t}$ encodes what information each token seeks, and $\begin{bmatrix}\boldsymbol{k}_n^{h,t}\; \boldsymbol{v}_n^{h,t}\end{bmatrix}$ encodes what information others provide. This motivates our reformulation of local self-attention computation in (\ref{reeq:update_local_o}), where we position KVs as conditioning variables and queries as optimization targets, mirroring FL's local model training conditioned on local datasets. From \textit{\textbf{{Observation}} \ref{O1}}, FedAttn approaches its centralized counterpart when local attention converges to global attention, i.e., block-diagonal attention pattern where inter-participant attention weights vanish as\begin{align}\boldsymbol{q}_n^{h,t}\left(\boldsymbol{k}_{n'}^{h,t}\right)^\top=\boldsymbol{0}_{L_{n} \times L_{n'}}\quad\forall n\neq n'.\end{align}\end{enumerate}Note that Queries, Keys and Values are initialized from hidden representations at the beginning of each local forward. This reveals that in contrast to FL's static data distribution, FedAttn exhibits \textbf{Dynamic Attention Distribution}\cite{DBLP:conf/nips/JiangLZWLAHA0L024,DBLP:journals/corr/abs-2110-11299} across local updates due to the following two interdependent factors.\begin{enumerate}[label=\bf\textit{\arabic*)}]
		\item Transformer blocks learn block-specific weight matrix $\begin{bmatrix}\boldsymbol{W}_{Q}^m\;\boldsymbol{W}_{K}^m\; \boldsymbol{W}_{V}^m\end{bmatrix}$ for Query-Key-Value projection, each reconstructing a specialized information retrieval objective within its respective contextual representational subspace.
		\item Hidden representations evolve tokens' contextual semantics progressively, reshaping token relevance across blocks.\end{enumerate}
\end{observation}

\begin{observation}[\textnormal{Aggregation method}]The key algorithmic insight of FedAttn lies in self-attention's limited scope to local input sequences, with periodically expanded scope to the global sequence every $H$ Transformer blocks, mirroring FL's approach where local updates extract specific patterns from local data distributions while periodic aggregation integrates global knowledge from the diverse datasets of all participants.\begin{enumerate}[label=\bf\textit{\arabic*)}]
	\item FL's model aggregation typically follows the federated averaging (FedAvg) algorithm with the averaging weight proportional to the local dataset size, i.e., \begin{align}\alpha_n=\frac{D_n}{\sum_{n=1}^N D_n}.\end{align}Under the IID data distribution, FedAvg achieves optimal convergence as it guarantees unbiased gradient estimation across all participants. Under non-IID settings, adaptive aggregation method reduces local model bias by prioritizing participants with globally representative data distributions, enhancing global model convergence.
	\item FedAttn aggregates local KV matrices by concatenating them into a global one. Given that attention distribution typically exhibits unbalanced relevance across tokens, \textbf{Sparse KV Exchange} method reduces \textit{a) Communication overhead} by having participants selectively transmit critical KVs, and \textit{b) Computational overhead} of global self-attention computations due to the reduced dimensionality of global KV matrix. This \textbf{Adaptive KV Aggregation} method can be formulated as\begin{align}\begin{bmatrix}{\boldsymbol{K}}^{h,t}\; {\boldsymbol{V}}^{h,t}\end{bmatrix}=\sum_{n=1}^{N}\boldsymbol{\pi}_n(t) \begin{bmatrix}\boldsymbol{k}_n^{h,t}\; \boldsymbol{v}_n^{h,t}\end{bmatrix},\end{align}where the weight $\boldsymbol{\pi}_n(t)\in \{0,1\}^{L \times L_n}$ is defined as\begin{align}
		\boldsymbol{\pi}_n(t) &= \left( \sum_{i \in \mathcal{L}_n'(t)} \boldsymbol{e}_i \boldsymbol{1}^\top \right) \odot \boldsymbol{\Pi}_n,\; \mathcal{L}_n'(t) \subset \mathcal{L}_n.
	\end{align}This communication-computation efficiency comes at the cost of introducing approximation errors to attention outputs, establishing a efficacy-efficiency trade-off in FedAttn. In the limiting case where $\boldsymbol{\pi}_n(t)=\mathbf{0}^{L \times L_n}$, i.e., $\mathcal{L}_n'(t)=\emptyset$, the $n$-th participant is completely excluded from the KV aggregation in the $t$-th communication round, which yields maximum communication-computation efficiency at the cost of maximum approximation errors.\end{enumerate} 

\end{observation}

\section{Error Analysis}\label{Error Analysis}
Prior to analyzing error propagation, we make the following assumptions regarding self-attention mechanisms and FFN transformations.
\begin{assumption}\label{assumption1}
	\textit{(}Lipschitz continuity\textit{).} Self-attention mechanisms and FFN transformations are Lipschitz continuous\cite{DBLP:conf/icml/DasoulasSV21}, i.e., for any $m\in\mathcal{M}$, and $\boldsymbol{X}$, $\boldsymbol{Y}\in\mathbb{R}^{L\times d}$, the following hold that\begin{align}\label{assumption1-1}
		&\left\Vert\text{Attention}\Big(\mathcal{P}^m_\text{QKV}\left(\boldsymbol{X}\right)\Big)-\text{Attention}\Big(\mathcal{P}^m_\text{QKV}\left(\boldsymbol{Y}\right)\Big)\right\Vert_F\nonumber\\&\leq \varrho_m\left\Vert\boldsymbol{X}-\boldsymbol{Y}\right\Vert_F,
	\end{align}and\begin{align}\label{assumption1-2}
		&\left\Vert\mathcal{F}^m\left(\boldsymbol{X}\right)-\mathcal{F}^m\left(\boldsymbol{Y}\right)\right\Vert_F\leq \theta_m\left\Vert\boldsymbol{X}-\boldsymbol{Y}\right\Vert_F,
	\end{align}where $\varrho_m>0$ and $\theta_m>0$ denote the Lipschitz constants of the self-attention and FFN sub-layers of the $m$-th Transformer block, respectively.
\end{assumption}
\begin{assumption}\label{assumption2}
	\textit{(}Bounded local variances\textit{).} For any $n\in\mathcal{N}$, $m\in\mathcal{M}$, and $\boldsymbol{X}\in\mathbb{R}^{L\times d}$, the deviation between local and global attention outputs is bounded in Frobenius norm, i.e.,\begin{align}\label{assumption2-1}
		&\Big\Vert \text{Attn}\left(\mathcal{P}^m_\text{Q}\left(\boldsymbol{\Pi}_n\boldsymbol{X}\right)\Big|\mathcal{P}^m_\text{K}\left(\boldsymbol{\Pi}_n\boldsymbol{X}\right),\mathcal{P}^m_\text{V}\left(\boldsymbol{\Pi}_n\boldsymbol{X}\right)\right)\nonumber\\&-\text{Attn}\left(\mathcal{P}^m_\text{Q}\left(\boldsymbol{\Pi}_n\boldsymbol{X}\right)\Big|\mathcal{P}^m_\text{K}\left(\boldsymbol{X}\right),\mathcal{P}^m_\text{V}\left(\boldsymbol{X}\right)\right)\!\Big\Vert_F\!\leq\!\sigma_n^m.\!
	\end{align}
\end{assumption}\textit{\textbf{Assumption} \ref{assumption1}} establishes Lipschitz continuity throughout the Transformer architecture, providing a bound on the dynamics of error propagation for hidden representations during forward passes. \textit{\textbf{Assumption} \ref{assumption2}} characterizes the attention distribution across participants at different blocks, with $\sigma_n^m$ measuring how much local attention deviates from the global view. 
	
\begin{theorem}\label{theorem_1}
	Given the total number of Transformer blocks $M = HT$, the approximation error between FedAttn and centralized attention (CenAttn for short) can be bounded under \textit{\textbf{Assumptions} \ref{assumption1}} and \textit{\ref{assumption2}}, as follows:\begin{align}\label{eafsgrthy}\left\Vert \boldsymbol{X}^{T}-\boldsymbol{X}^*\right\Vert_F&\leq\sum_{t=0}^{T-1}\sum_{h=1}^{H-1} \underbrace{\left(\underbrace{\left(1+\theta_{Ht+h}\right)}_{\textit{(a.2)}} \underbrace{{\sum}_{n=1}^N \sigma_{{Ht+h},n}}_{\textit{(a.1)}}\right)}_{{\textit{(a) local self-attention deviation}}}\nonumber\\&\quad\times\underbrace{\left(\prod_{i=h+1}^{H} \left(1+\theta_{Ht+i}\right) \left(1+\varrho_{Ht+i}\right)\right)}_{\textit{(b) intra-round amplification}}\nonumber\\&\quad \times\underbrace{\left(\prod_{j=t+1}^{T-1} \prod_{i=1}^{H} \left(1+\theta_{Hj+i}\right) \left(1+\varrho_{Hj+i}\right)\right)}_{\textit{(c) inter-round amplification}}\end{align}
\end{theorem}\begin{IEEEproof}Please see the proof in Appendix A.\end{IEEEproof}
	
\textit{\textbf{Theorem} \ref{theorem_1}} reveals that the approximation error between FedAttn and CenAttn accumulates across local forwards and communication rounds, with each deviation term arising from local attention computation and its amplification through subsequent blocks. The key components are interpreted as follows.

\begin{remark}[\textnormal{Lipschitz gain}] 
	Given the Lipschitz constants of the self-attention mechanisms and FFN transformations $\theta_m,\varrho_m>0$, the Lipschitz gain of the $m$-th block yields\begin{equation}\gamma_m=\left(\theta_m+1\right) \left(\varrho_m+1\right)>1,\end{equation}where the ``$+1$" terms correspond to the identity mappings in the residual connections surrounding the self-attention and FFN sub-layers. 
\end{remark}

\begin{remark}[\textnormal{Error Injection}]
	Term \textit{(a)} captures the approximation error injected by local self-attention computation at the $m$-th Transformer block, i.e., during the $h$-th local forward of the $t$-th communication round.  
	\begin{enumerate}[label=\bf\textit{\arabic*)}]
		\item \textit{Self-attention deviation.} Term \textit{(a.1)} quantifies the deviation arising from local self-attention computation, where queries attend exclusively to local KV pairs rather than the global ones. This deviation formalizes the inter-participant attention distribution dynamics across Transformer blocks. Specifically, each block learns specialized Query-Key-Value projection matrices, refining hidden representations by analyzing contextual relevance between tokens with specialized representational subspaces. 
		\item \textit{FFN amplification.} The multiplicative factor $\left(1+\theta_{Ht+h}\right)$ quantifies how the self-attention deviation undergoes further amplification via the subsequent FFN transformation.\end{enumerate} 
	
	The summation of Term \textit{(a)} over $h \in \{1, 2, \ldots, H-1\}$ accumulates approximation errors within each communication round, excluding the $H$-th local forward where global self-attention computation over aggregated KV pairs operates without any error injection.
\end{remark}
	
\begin{remark}[\textnormal{Error Propagation}] 
	Terms \textit{(b)} and \textit{(c)} quantify intra-round and inter-round error amplification through subsequent Transformer blocks, respectively. Under \textit{\textbf{Assumptions} \ref{assumption1}} and \textit{\ref{assumption2}}, both terms exhibit multiplicative structure, reflecting the compositional nature of deep architectures where approximation errors propagate and amplify sequentially through self-attention and FFN operations according to their respective Lipschitz gains. 
	
	This echo chamber phenomenon indicates that earlier local forwards dominate the error landscape. Intuitively, errors injected at earlier local forwards traverse more subsequent layers, accumulating larger amplification factors and rendering them disproportionately impact on final outputs.
\end{remark}

%From \textbf{Theorem} \ref{theorem_1}, we establish \textbf{Corollary} \ref{lemma_4} under uniform Lipschitz constants to investigate the approximation error of FedAttn scaling with the number of Transformer blocks $M$ and local forwards $H$ as follows.
From \textit{\textbf{Theorem} \ref{theorem_1}}, we establish \textit{\textbf{Corollary} \ref{lemma_4}} under uniform Lipschitz constants to investigate the approximation error of FedAttn scaling with the number of local forwards as follows.

\begin{corollary}\label{lemma_4}Under \textit{\textbf{Assumptions} \ref{assumption1}} and \textit{\ref{assumption2}}, suppose $\theta_m\leq \theta$, $\varrho_m\leq \varrho$, and $\sigma_n^m\leq\sigma_n$ hold for all $m\in\mathcal{M}$, and denote $\gamma=\left(1+\theta\right) \left(1+\varrho\right)$. Given the total number of Transformer blocks $M = HT$, the approximation error between FedAttn and CenAttn can be bounded as follows:\begin{align}\label{eafsgrthrthy}\left\Vert \boldsymbol{X}^{T}-\boldsymbol{X}^*\right\Vert_F\leq\left(\left(1+\theta\right) \sum_{n=1}^N \sigma_n\right)\underbrace{\frac{\gamma^M-1}{\gamma-1}}_{\textit{(d)}}\underbrace{\left(1-\frac{\gamma-1}{\gamma^H-1}\right)}_{\textit{(e)}}.\end{align}\end{corollary}\begin{IEEEproof}Please see the proof in Appendix B.\end{IEEEproof}

\begin{remark}
	Term \textit{(e)} increases monotonically from $0$ at $H=1$ to $1$ as $H\to\infty$, characterizing FedAttn's trade-off between communication efficiency and approximation errors. Specifically, when $H=1$, each Transformer block performs global attention computation over aggregated KV pairs without introducing any approximation errors. When $H=M$, each individual block is restricted to local attention computation, reducing FedAttn to a fully local attention (LocAttn for short) mechanism and yielding entirely local LLM inference.
\end{remark}

\begin{remark}[\textnormal{Diminishing gains in communication efficiency}]\label{Remark:Diminishing gains in communication efficiency}The Taylor series expansion of term \textit{(e)} around $\gamma = 1 + \varepsilon$ with $\varepsilon \to 0$ gives\begin{align}
		1-\frac{\gamma-1}{\gamma^H-1} &= 1-\frac{\gamma-1}{H \varepsilon + \frac{H}{2}(H-1) \varepsilon^2 + \mathcal{O}\left(\varepsilon^3\right)} \nonumber\\&= 1 - \frac{1}{H} + \mathcal{O}\left(\gamma - 1\right),\end{align}which establishes that, for any fixed number of local forwards $H$, FedAttn's approximation error asymptotically approaches the limit $1 - \frac{1}{H}$ as $\gamma\to 1$. The marginal reduction in communication overhead when increasing $H$ to $H+1$ is\begin{align}\frac{1}{H} - \frac{1}{H+1} = \frac{1}{H(H+1)},\end{align}with the marginal increase in approximation error being of the same magnitude. This reveals that small $H$ exhibits substantial marginal effects in both communication overhead and approximation error. Conversely, in the large $H$ regime, each additional local forward pass yields progressively diminishing gains in FedAttn's communication efficiency, while performance degradation increasingly intensifies due to accumulated approximation errors.
\end{remark}\begin{IEEEproof}Please see the details in Appendix C.\end{IEEEproof}

\begin{theorem}\label{theorem_3}
	Assuming a variable number of local forwards $H_t>0$ across communication rounds satisfying  $\sum_{t=0}^{T-1} H_t= M$, the approximation error between FedAttn and CenAttn can be bounded under \textit{\textbf{Assumptions} \ref{assumption1}} and \textit{\ref{assumption2}}, as follows:\begin{align}\label{sefrdgthyfjfd}&\left\Vert \boldsymbol{X}^{T}-\boldsymbol{X}^*\right\Vert_F\leq\sum_{m=0}^{M-1}\underbrace{\left(\left(1+\theta_{m}\right) \sum_{n=1}^N \sigma_n^m\right)}_{\textit{(f.1)}}\underbrace{\left(\prod_{i=m+1}^{M-1} \left(1+\theta_{i}\right) \Big(1+\varrho_{i}\Big)\right)}_{\textit{(f.2)}}\nonumber\\&-\sum_{t=0}^{T-1}\underbrace{\left(\left(1+\theta_{\sum\limits_{j=0}^{t-1}H_j}\right)\sum_{n=1}^N \sigma_n^{\sum\limits_{j=0}^{t-1}H_j}\right)}_{\textit{(g)}}\underbrace{\left(\prod_{m=\sum\limits_{j=0}^{t-1}H_j+1}^{M-1} \left(1+\theta_{m}\right) \left(1+\varrho_{m}\right)\right)}_{\textit{(h)}}.\end{align}\end{theorem}\begin{IEEEproof}Please see the proof in Appendix D.\end{IEEEproof}
	
\begin{remark}[\textnormal{Where to Perform Global Attentions}]\label{remark:where to place global attentions}Theorem \ref{theorem_3} decomposes the approximation error of FedAttn regarding determining which Transformer blocks perform global attention, yielding key insights as follows:\begin{enumerate}[label=\bf\textit{\arabic*)}]
		\item Term \textit{(f)} represents FedAttn's approximation error when implementing fully local computation, i.e., when $H=M$. Each Transformer block introduces a deviation, as shown in term \textit{(f.1)}, which is then amplified by all subsequent Lipschitz gains, as shown in term \textit{(f.2)}.
		\item Terms \textit{(g)} and \textit{(h)} quantify the reduction in approximation error achieved by performing global attention computation at the $m$-th block. Term \textit{(g)} captures the deviation between the local and global self-attention outputs at that block, and term \textit{(h)} corresponds to the amplification factor of this deviation through subsequent blocks. This error reduction at the $m$-th block can be rewritten as\begin{align}
			\Gamma_m = (1+\theta_{m})\sum_{n=1}^N\sigma_n^m \prod_{i=m+1}^{M-1}(1+\theta_i)(1+\varrho_i),\!\end{align}impling that shallower Transformer blocks and those with higher deviation $\sigma_n^m$ that indicates stronger inter-participant attention imbalance yield greater error reductions, making them more effective for performing global self-attention.
	\end{enumerate} 
\end{remark}

\section{Experimental Results}\label{Experimental Results}

In this section, we investigate two key questions through experiments: \textit{1) {How effective and efficient is FedAttn?}} We evaluate FedAttn across varying numbers of local forwards and participants, examining the trade-offs between response quality and communication/computational cost. \textit{2) What mechanisms underpin FedAttn?} We investigate two critical aspects of \textit{a) error propagation dynamics}, and\textit{ b) attention distribution} both within and across participants. 

To further improve FedAttn, we explore optimization opportunities via Sparse Attention Mechanism: \textit{1) adaptive KV aggregation}, \textit{2) sparse local attention}, \textit{3) sparse KV exchange}.

\subsection{Setup}
\subsubsection{Backbones and Datasets} 
In the following experiments, we evaluate FedAttn on the Qwen2.5 family of base models in sizes of 0.5B, 1.5B, 3B, and 7B parameters, using Grade School Math 8K (GSM8K) for mathematical reasoning tasks.

\subsubsection{Input Format} 
\begin{figure}[!t]
	\centering\
	\includegraphics[width=3.5in,angle=0]{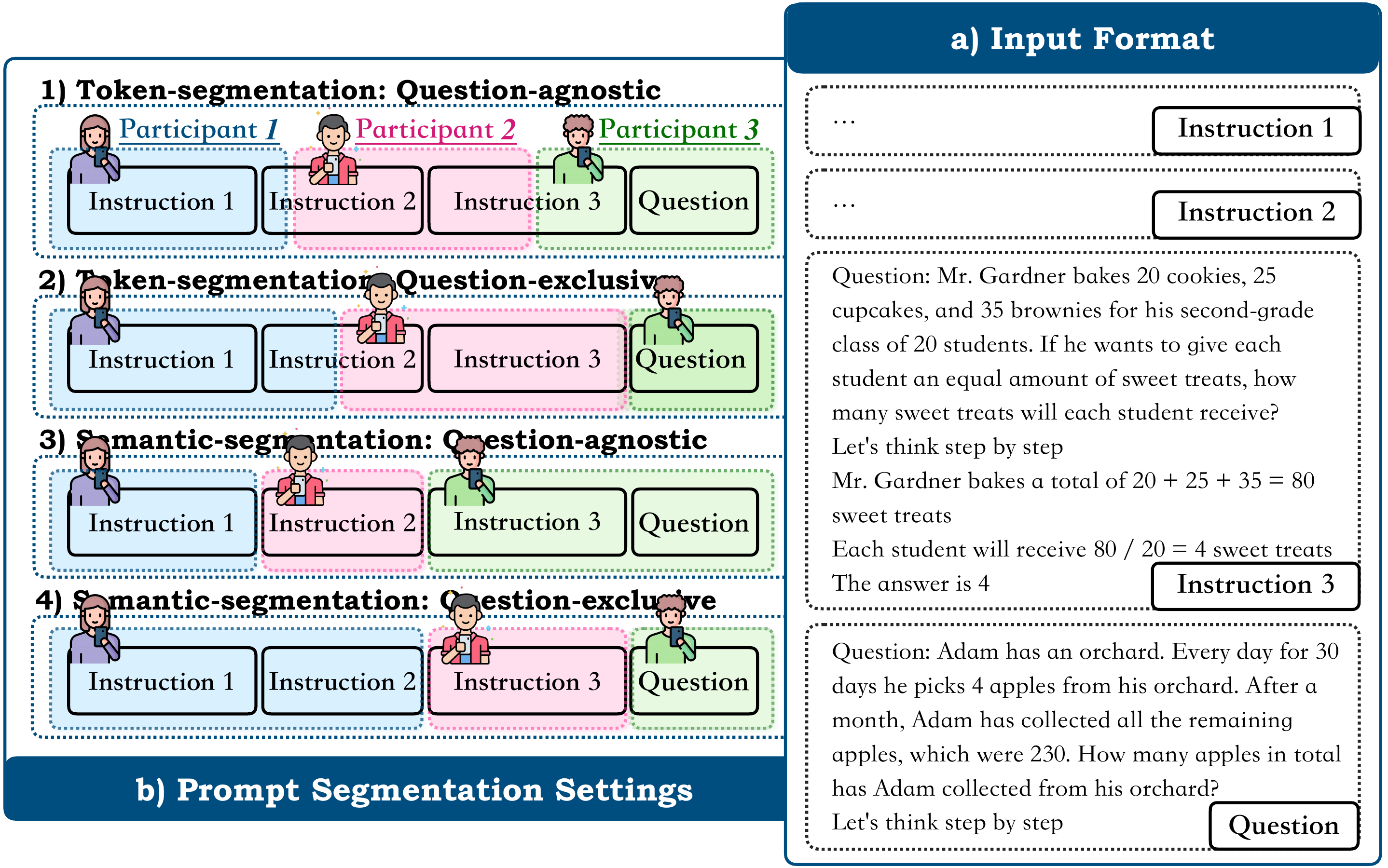}
	\caption{Illustration of\textit{ a) Input Format} and \textit{b) Input Segmentation Settings}. We exemplify the input format with a 3-shot instructional examples and demonstrate 4 segmentation settings under 3 participants, each holding a segment as its local input sequence.}
	\label{fig2}
\end{figure}
We evaluate FedAttn on GSM8K's test split under CoT prompting with Qwen's official evaluation template, utilizing few-shot examples that demonstrate step-by-step mathematical reasoning followed by the target question. We designate the $N$-th participant holding the target question as the task publisher.

To evaluate FedAttn across varying attention distribution within and across participants, we consider four input segmentation settings: \textit{a) Tok-seg: Q-ag}. Partition the global input sequence uniformly by token count across all participants. \textit{b) Tok-seg: Q-ex}. Allocate the complete question to the $N$-th participant, and uniformly partition examples by token count among others. \textit{c) Sem-seg: Q-ag}. Segment the global sequence at semantic boundaries, and uniformly distribute units across all participants. \textit{d) Sem-seg: Q-ex}. Allocate the question to the $N$-th participant, and distribute examples intact among others.

These segmentation settings form a $2\times 2$ grid along two orthogonal dimensions: \textit{a) Token-segmentation vs. Semantic-segmentation}. Token-segmentation maintains uniform communication and computational costs across participants but risks fragmenting semantic boundaries within instruction examples and the target question. In contrast, semantic-segmentation maintains the integrity of both instruction examples and the target question, preserving contextual dependencies within each semantic unit accessible to local attention computation. \textit{b) Question-agnostic vs. Question-exclusive}. The target question defines the objective of the inference task, representing the most semantically critical component of the global input sequence. Question-agnostic segmentation distributes the target question alongside instructional examples across participants, providing local access to both task objectives and instructions. Question-exclusive segmentation involves isolating the target question within a single participant while dispersing instructional examples, rendering the target question dependent on KV exchange among participants for accessing instructions during FedAttn inference.

\subsubsection{Evaluation Metrics} 

Although in practical deployment only the task publisher generates the final response, we have each participant generate a response in our experiments to evaluate how cross-participant attention distribution affects FedAttn performance. Specifically, each participant maintains KV caches in memory from Prefilling, and reuses them at decoding steps. Following prior practice, we report Pass@1 Exact Match (EM) accuracy on GSM8K for each individual participant to quantify the response quality of FedAttn.

We access the communication and computational efficiency of FedAttn as follows: \textit{a) Communication cost} is measured as average bits transmitted per participant for KV exchanges during Prefilling. \textit{b) Computational cost} encompasses average FLOPs and peak memory usage per participant during both Prefilling and Decoding.
\subsection{Results and Discussion}
\subsubsection{\textbf{Trade-off Between Response Quality and Communication Cost}}
\begin{figure*}[!t]
	\centering
	\begin{subfigure}[!t]{\textwidth}
		\includegraphics[width=\textwidth]{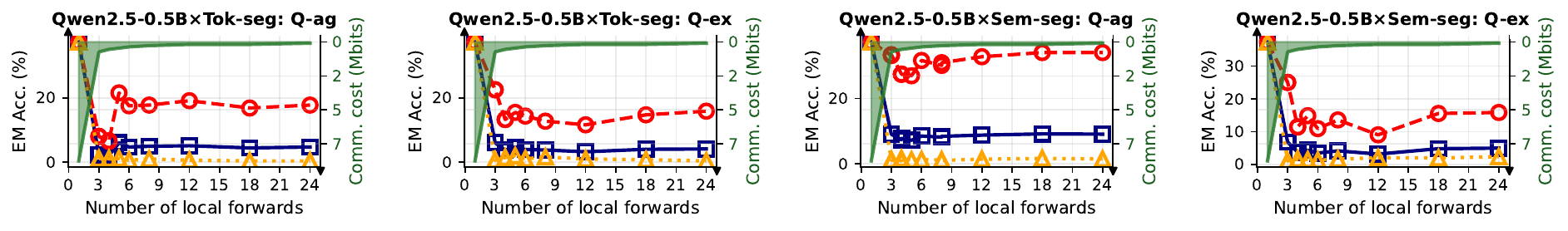}
		\label{fig_exp_1:1}
	\end{subfigure}\vspace{-12pt}
	\begin{subfigure}[!t]{\textwidth}
		\includegraphics[width=\textwidth]{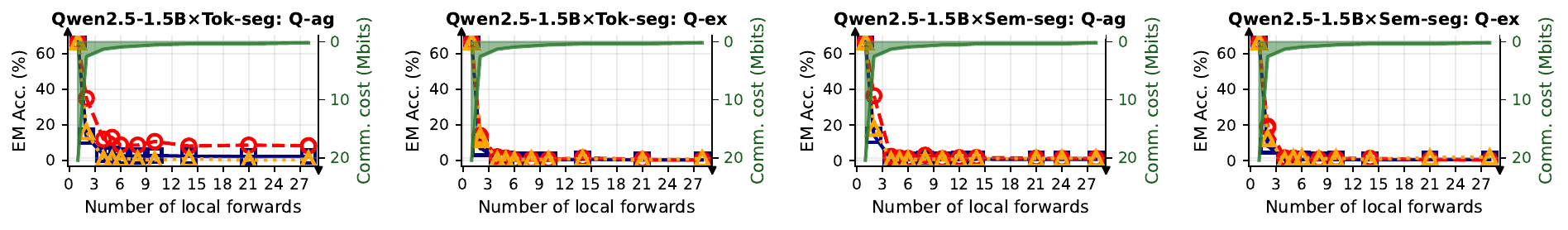}
		\label{fig_exp_1:2}
	\end{subfigure}\vspace{-12pt}
	\begin{subfigure}[!t]{\textwidth}
		\includegraphics[width=\textwidth]{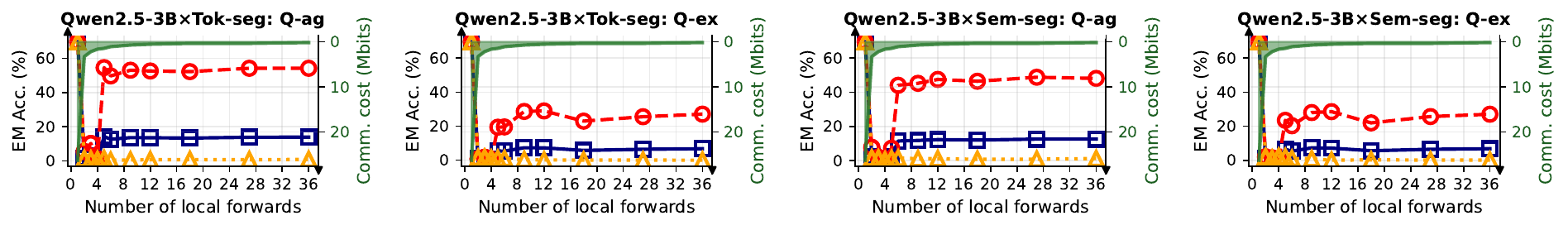}
		\label{fig_exp_1:3}
	\end{subfigure}\vspace{-12pt}
	\begin{subfigure}[!t]{\textwidth}
		\includegraphics[width=\textwidth]{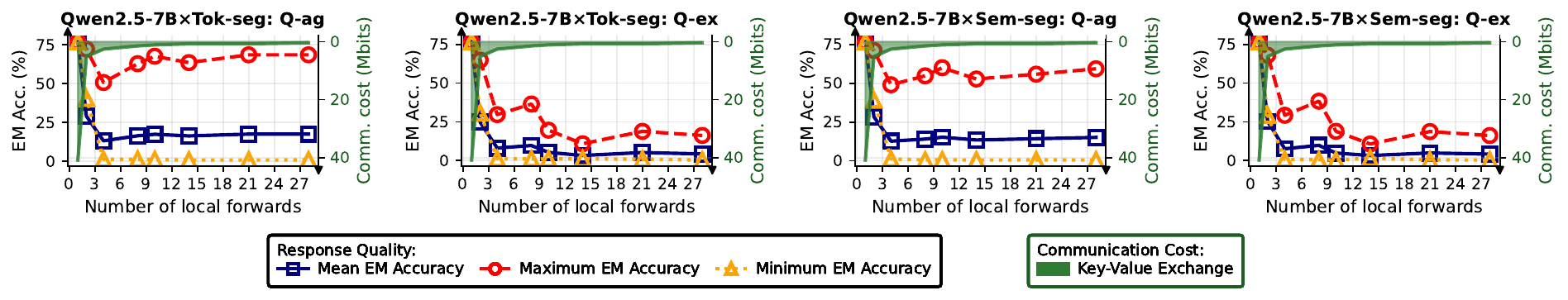}
		\label{fig_exp_1:4}\vspace{-10pt}
	\end{subfigure}	
	\caption{Trade-off between response quality and communication cost. Evaluated with 4-shot prompting, greedy decoding, max 256 new tokens.}
	\label{fig_exp_1}
\end{figure*}
Fig.~\ref{fig_exp_1} illustrates FedAttn's efficacy-efficiency trade-off across varying local forwards, with EM accuracy on the primary axis and communication cost on the secondary. We organize results by model size in rows and input segmentation in columns, and report the mean EM accuracy along with the minimum and maximum values across all participants to capture the variation in response quality. The number of local forwards ranges from $H=1$ to $H=M$, where $H=1$ reduces FedAttn to CenAttn while $H=M$ reduces to LocAttn. The key observations are as follows.\begin{enumerate}[label=\bf\textit{\alph*)}]
	\item EM accuracy decreases with $H$ while communication cost diminishes correspondingly, exhibiting \textbf{Diminishing Returns} that both metrics decrease sharply at small $H$ before plateauing as $H$ increases. This aligns with \textit{\textbf{Remark}~\ref{Remark:Diminishing gains in communication efficiency}} where marginal reduction in communication cost diminishes as $\mathcal{O}\left({1}/{H^2}\right)$, and approximation error accumulates as $\mathcal{O}\left({1}/{H}\right)$, and indicates small $H$ captures most communication savings with limited response quality degradation. 
	\item Larger models exhibit \textbf{Robustness to Reduced Global Self-Attention}, maintaining higher accuracy and slower degradation via representational redundancy. For instance, 7B retains relatively high response quality even at large $H$, while 0.5/1.5B depend heavily on KV exchange to access the complete contextual information. 
	\item \textbf{Segmentation Hierarchy} persists: Question-agnostic outperforms Question-exclusive by enabling local access to both questions and instructions. Token-segmentation suits small models (0.5B) with balanced segment lengths, while Semantic-segmentation suits large models by preserving semantic boundaries. This is due to the fact that small LLMs suffer from short segments lacking sufficient local contexts while long segments exceeding narrow attention window and accumulating noise. 
	\item \textbf{Performance Drops} occur at intermediate $H$ where synchronization frequency of KV exchange induces conflicts between local independence and global coherence. Critically, 3B exhibits pronounced valley, possessing sufficient capacity to utilize remote context yet lacking robustness against stale information. 
	\item Minimum EM accuracy deteriorates more rapidly than the maximum, reflecting \textbf{Inter-participant Attention Distribution}: participants with target questions or key instructions remain robust, while those dominated by auxiliary tokens degrade sharply with reduced global attention. Additionally, Semantic-segmentation and Question-agnostic reduce performance divergence by preserving semantic integrity.\end{enumerate}

\subsubsection{\textbf{Trade-off Between Response Quality and Computational Cost}}
\begin{figure*}[!t]
	\centering
	\begin{subfigure}[!t]{\textwidth}
		\includegraphics[width=\textwidth]{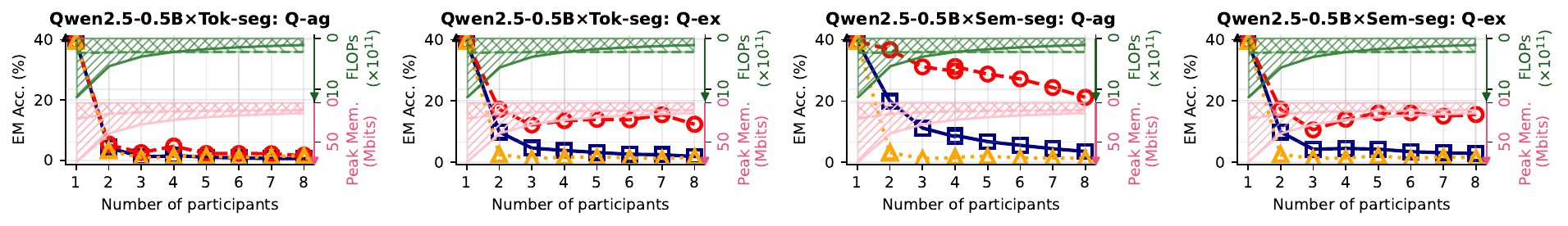}
		\label{fig_exp_2:1}
	\end{subfigure}\vspace{-12pt}
	\begin{subfigure}[!t]{\textwidth}
		\includegraphics[width=\textwidth]{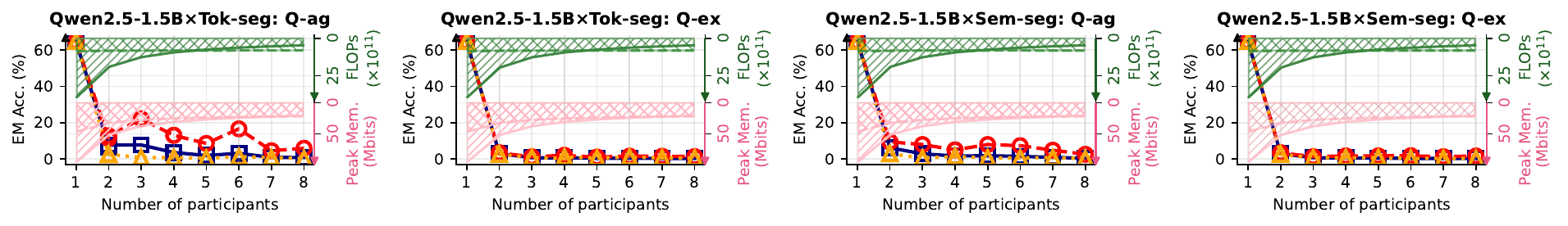}
		\label{fig_exp_2:2}
	\end{subfigure}\vspace{-12pt}
	\begin{subfigure}[!t]{\textwidth}
		\includegraphics[width=\textwidth]{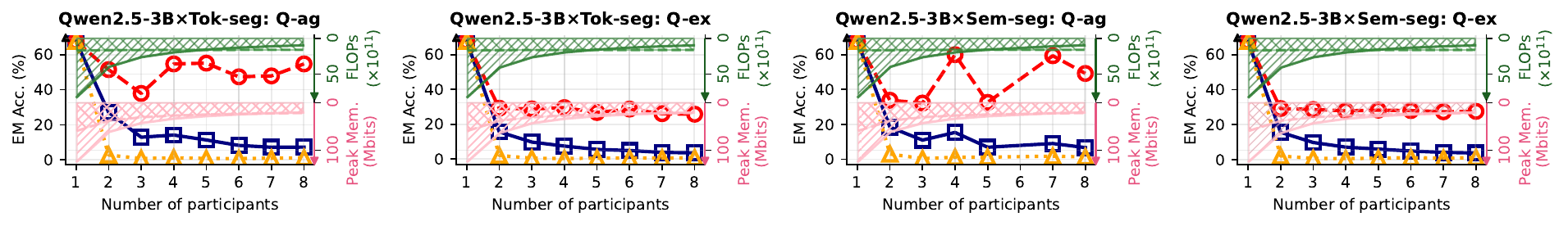}
		\label{fig_exp_2:3}
	\end{subfigure}\vspace{-12pt}
	\begin{subfigure}[!t]{\textwidth}
		\includegraphics[width=\textwidth]{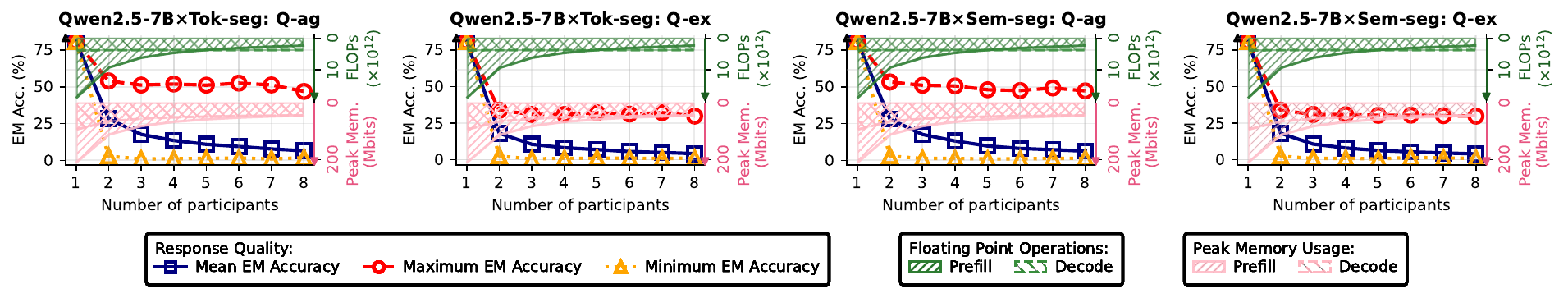}
		\label{fig_exp_2:4}\vspace{-10pt}
	\end{subfigure}	
	\caption{Trade-off between response quality and computational cost. Evaluated with 8-shot prompting, greedy decoding, max 256 new tokens.}
	\label{fig_exp_2}
\end{figure*}

Fig.~\ref{fig_exp_2} illustrates the efficacy-efficiency trade-off across varying numbers of participants, where the upper panel shows FLOPs and the lower panel shows peak memory. The number of local forwards ranges from $N=1$ to the number of few-shot examples. When $N=1$, FedAttn reduces to CenAttn. The key observations are as below:\begin{enumerate}[label=\bf\textit{\alph*)}]
	\item EM accuracy decreases with $N$ while computational costs diminish correspondingly. FLOPs and peak memory decrease approximately quadratically at Prefilling and linearly at Decoding, consistent with the theory. 
	\item Large models (3B/7B) exhibit attenuated accuracy decay thanks to representational redundancy. Semantic-segmentation and Question-agnostic demonstrate greater resilience due to local access to both questions and instructions and the integrity of semantic units.\end{enumerate}

\subsubsection{\textbf{Error Propagation} (Where to perform global attention)}

\begin{figure}[!t]
	\centering
	\includegraphics[width=3.5in,angle=0]{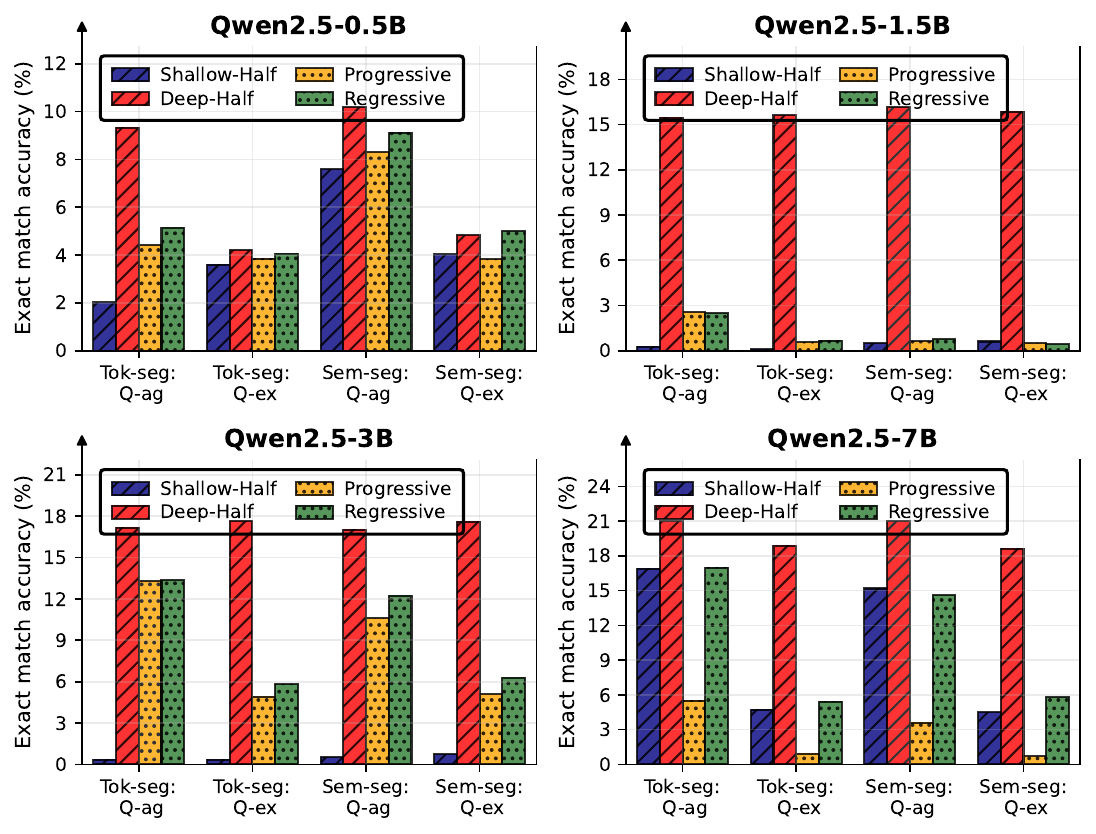}
	\caption{Response quality under 4 synchronization schemes for KV exchange. Evaluated with 4 participants, 4 communication rounds, 4-shot prompting, greedy decoding, max 256 new tokens.}
	\label{fig_exp_3}
\end{figure}

Fig.~\ref{fig_exp_3} compares four alternative synchronization schemes for KV exchange: \textit{a) Shallow-Half} and \textit{b) Deep-Half}, concentrating KV exchanges in shallower and deeper halves, \textit{c) Progressive} and \textit{d) Regressive}, with synchronization intervals increasing and decreasing with depth. Across all model sizes and input segmentations, Deep-Half substantially outperforms Shallow-Half, and Regressive outperforms Progressive, revealing that KV exchanges at deeper blocks are significantly more effective for maintaining response quality of FedAttn.

This experimental finding contradicts \textit{\textbf{Theorem} \ref{theorem_3}}, which predicts early synchronization should be more effective by immediately correcting deviations between local and global self-attention and thereby preventing subsequent error propagation. This discrepancy reveals critical insights into FedAttn's error propagation dynamics: \begin{enumerate}[label=\bf\textit{\alph*)}]
	\item\textit{Architectural mechanisms} such as residual connections, layer normalization, and multi-head redundancy substantially attenuate error propagation from shallow blocks, reducing early synchronization benefits. 
	\item Self-attention deviations $\sigma_n^m$ increase significantly with depth as deeper blocks produce highly abstract representations wherein $\sigma_n^m$ encodes substantial semantic correction information, further magnified by progressively sparse attention patterns.\end{enumerate}

\subsubsection{\textbf{Adaptive KV Aggregation} (Which participant contribu-tes most)}

\begin{figure}[!t]
	\centering
\includegraphics[width=3.5in,angle=0]{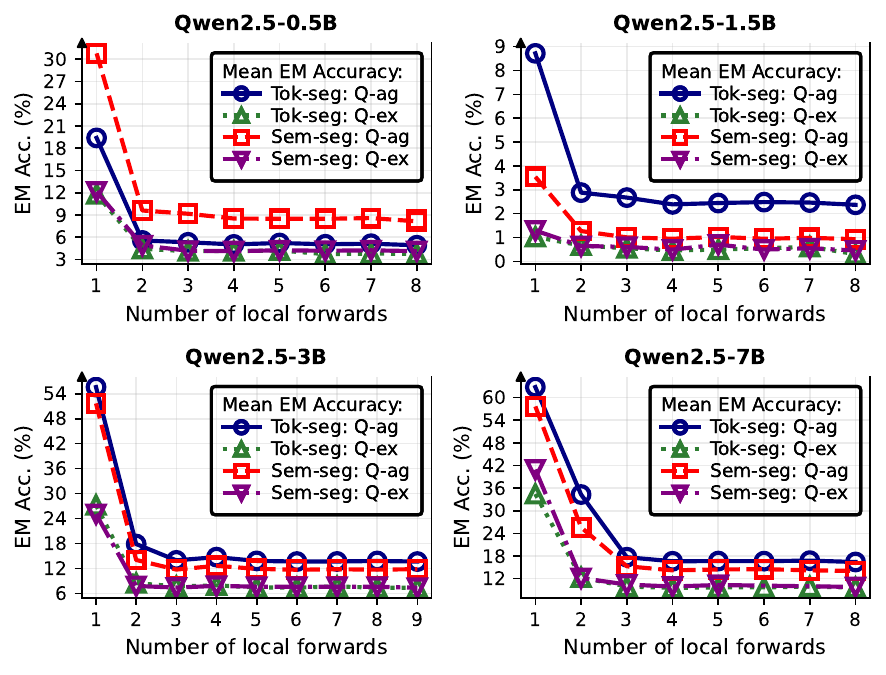}
\caption{Response quality under varying synchronization intervals for task publisher with X-axis denoting its local forwards and others fixed at 8 (0.5B/1.5B/7B) or 9 (3B), 4 participants, 4-shot prompting, greedy decoding, max 256 new tokens.}
\label{fig_exp_4}
\end{figure}

Fig.~\ref{fig_exp_4} shows EM accuracy increases monotonically with the synchronization frequency of the task publisher.\begin{enumerate}[label=\bf\textit{\alph*)}] 
	\item Large models such as 7B achieve substantial improvements by effectively leveraging enriched context, while small models such as 0.5B plateau rapidly due to limited model capacity. 
	\item Semantic segmentation and Question-agnostic consistently outperform Token segmentation and Question-exclusive by preserving semantic integrity and providing local access to both the question and instructions.\end{enumerate}These findings indicate that increasing synchronization frequency for critical participants enables more informative contextual contributions, improving the overall response quality.

\subsubsection{\textbf{Sparse Local Attention} (How many tokens to prefill)}
\begin{figure}[!t]
	\centering
\includegraphics[width=3.5in,angle=0]{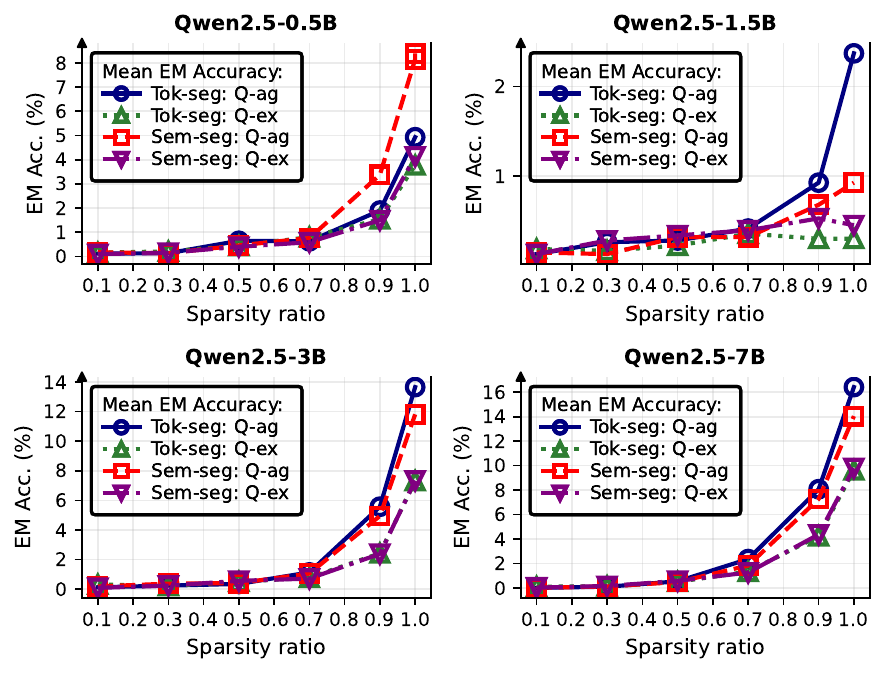}
\caption{Response quality under varying token sparsity ratios for local self-attention, 4 participants, 4 communication rounds, 4-shot prompting, greedy decoding, max 256 new tokens.}
\label{fig_exp_5}
\end{figure}
Fig.~\ref{fig_exp_5} plots the mean EM accuracy under different sparsity ratios for local self-attention, wherein participants randomly sample the local input tokens before inference tasks. Across all model sizes and input segmentations, EM accuracy decreases with decreasing sparsity ratio, revealing that random sparsification reduces computational cost by constraining the attention scope at the expense of response quality. 

Larger models such as 7B show substantial robustness to sparse local self-attention, maintaining relatively high EM accuracy at moderate sparsity ratios thanks to representational redundancy that compensates for the reduced local attention scope. Token-segmentation and Question-exclusive consistently underperform Semantic segmentation and Question-agnostic, exhibiting lower tolerance to sparse attention caused by fragmenting semantic boundaries and isolating questions from instructions.

\subsubsection{\textbf{Sparse KV Exchange} (How many KVs to exchange)}
\begin{figure}[!t]
	\centering
	\includegraphics[width=3.5in,angle=0]{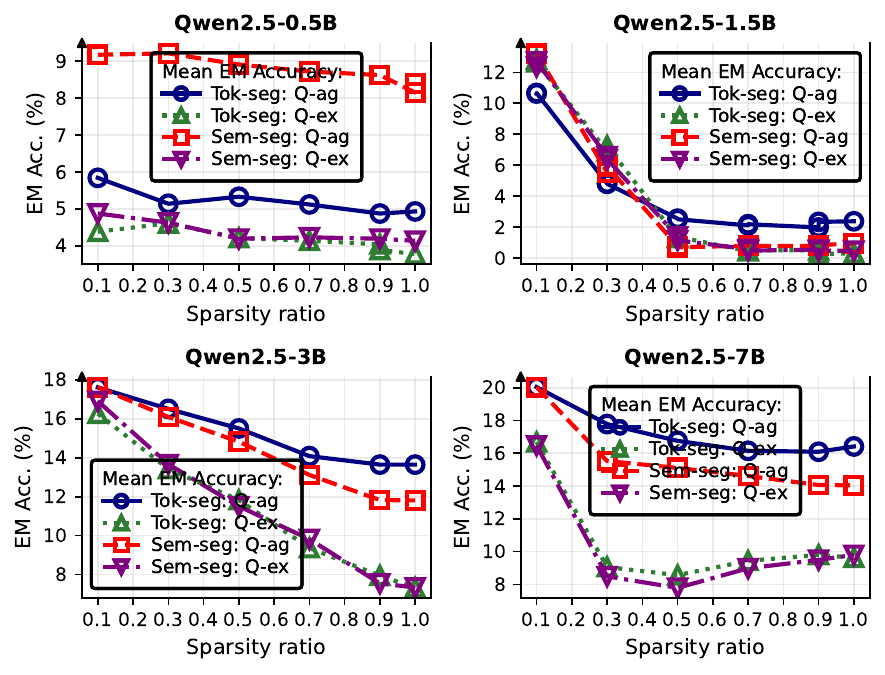}
	\caption{Response quality under varying sparsity ratios for KV exchange, 4 participants, 4 communication rounds, 4-shot prompting, greedy decoding, max 256 new tokens.}
	\label{fig_exp_6}
\end{figure}
Fig.~\ref{fig_exp_6} plots the mean EM accuracy under different sparsity ratios for KV exchange, where participants exchange randomly sampled KV subsets during each communication round. Across all model sizes and input segmentations, EM accuracy increases with increasing sparsity, which shows that sparse KV exchange improves response quality while reducing communication overhead. This finding challenges the intuition that self-attention over all input tokens maximizes response quality, stemming from key mechanisms:\begin{enumerate}[label=\bf\textit{\alph*)}] 
	\item Random sparsification acts as a regularizer, filtering \textbf{Temporal Staleness} and semantic misalignment from remote KV pairs that inject conflicting contextual information through attention mechanisms and impede token representation refinement across layers. 
	\item Noisy KV pairs disperse attention, where softmax normalization disperses weights across tokens and diminishes critical information. Sparse attention mitigates this dilution by limiting attention scope, reducing attention entropy and sharpening focus on semantically critical tokens.\end{enumerate}

A critical distinction emerges between sparse local attention and KV exchange as follows:\begin{enumerate}[label=\bf\textit{\alph*)}]
	\item\textit{Sparse local attention} discards input tokens prior to self-attention, causing irreversible information loss and thereby monotonic accuracy degradation. 
	\item\textit{Sparse KV exchange} preserves self-attention on complete local tokens within each participant while sparsifying remote KVs during global self-attention, effectively mitigating noisy, stale, and semantically misaligned cross-participant information while preserving local contextual integrity.\end{enumerate}

\section{Conclusion}\label{Conclusion}

In this work, we have proposed FedAttn, a new distributed self-attention paradigm that enables multiple participants to collaboratively generate LLM responses without exposing private prompts. We have conducted a theoretical analysis to characterize the error propagation dynamics of FedAttn and the fundamental trade-off between response quality and communication efficiency. Extensive experiments on the Qwen 2.5 model family and the GSM8K benchmark have verified the theoretical analysis, clearly demonstrating \textit{1)} trade-offs between response quality and communication/computational cost across varying numbers of local forwards and participants, \textit{2)} intensified error accumulation in deeper Transformer layers and imbalanced attention distributions both within and across participants, and \textit{3)} enhanced efficacy and efficiency through sparse attention and adaptive KV aggregation.

By advancing this federated paradigm for collaborative LLM inference at the edge, we aim to shift the focus of LLM research from an exclusive emphasis on model performance and towards distributed inference methodologies that address the limitations of on-device resources while adhering to strict privacy requirements, ultimately unlocking the transformative potential of LLMs in real-world edge networks.

\ifCLASSOPTIONcaptionsoff
\newpage
\fi
\bibliographystyle{IEEEtran}
\bibliography{references}

\begin{thebibliography}{10}
\providecommand{\url}[1]{#1}
\csname url@samestyle\endcsname
\providecommand{\newblock}{\relax}
\providecommand{\bibinfo}[2]{#2}
\providecommand{\BIBentrySTDinterwordspacing}{\spaceskip=0pt\relax}
\providecommand{\BIBentryALTinterwordstretchfactor}{4}
\providecommand{\BIBentryALTinterwordspacing}{\spaceskip=\fontdimen2\font plus
\BIBentryALTinterwordstretchfactor\fontdimen3\font minus
  \fontdimen4\font\relax}
\providecommand{\BIBforeignlanguage}[2]{{%
\expandafter\ifx\csname l@#1\endcsname\relax
\typeout{** WARNING: IEEEtran.bst: No hyphenation pattern has been}%
\typeout{** loaded for the language `#1'. Using the pattern for}%
\typeout{** the default language instead.}%
\else
\language=\csname l@#1\endcsname
\fi
#2}}
\providecommand{\BIBdecl}{\relax}
\BIBdecl

\bibitem{DBLP:journals/tai/HagosBR24}
D.~H. Hagos, R.~Battle, and D.~B. Rawat, ``Recent advances in generative {AI}
  and large language models: Current status, challenges, and perspectives,''
  \emph{{IEEE} Trans. Artif. Intell.}, vol.~5, no.~12, pp. 5873--5893, 2024.

\bibitem{DBLP:conf/icaiic/Saha24}
B.~K. Saha, ``Generative artificial intelligence for industry: Opportunities,
  challenges, and impact,'' in \emph{Proc. Int. Conf. Artif. Intell. Inf.
  Commun., \textnormal{Osaka, Japan, Feb. 19-22, 2024}}, pp. 81--86.

\bibitem{DBLP:conf/hpec/Li0GT24}
B.~Li, Y.~Jiang, V.~Gadepally, and D.~Tiwari, ``{LLM} inference serving: Survey
  of recent advances and opportunities,'' in \emph{Proc. IEEE High Perform.
  Extreme Comput. Conf., \textnormal{Wakefield, MA, USA, Sep. 23-27, 2024}},
  pp. 1--8.

\bibitem{11177548}
W.~Wang, K.~Li, B.~Ji, X.~Liu, J.~Yu, and Q.~Wu, ``A survey of ai inference
  technologies for on-device systems,'' \emph{IEEE Internet Things J.}, pp.
  1--1, 2025.

\bibitem{DBLP:conf/infocom/MaGZ025}
M.~Ma, C.~Gong, L.~Zeng, and Y.~Yang, ``Multi-tier multi-node scheduling of
  {LLM} for collaborative {AI} computing,'' in \emph{Proc. IEEE Conf. Comput.
  Commun., \textnormal{London, U.K., May 19-22, 2025}}, pp. 1--10.

\bibitem{DBLP:journals/corr/abs-2507-16731}
S.~Li, H.~Wang, W.~Xu, R.~Zhang, S.~Guo, J.~Yuan, X.~Zhong, T.~Zhang, and
  R.~Li, ``Collaborative inference and learning between edge {SLMs} and cloud
  {LLMs}: {A} survey of algorithms, execution, and open challenges,''
  \emph{CoRR}, vol. abs/2507.16731, 2025.

\bibitem{10835069}
G.~Qu, Q.~Chen, W.~Wei, Z.~Lin, X.~Chen, and K.~Huang, ``Mobile edge
  intelligence for large language models: A contemporary survey,'' \emph{{IEEE}
  Commun. Surv. Tutorials}, pp. 1--1, 2025.

\bibitem{DBLP:journals/csur/ZhengCQSSC25}
Y.~Zheng, Y.~Chen, B.~Qian, X.~Shi, Y.~Shu, and J.~Chen, ``A review on edge
  large language models: Design, execution, and applications,'' \emph{{ACM}
  Comput. Surv.}, vol.~57, no.~8, pp. 209:1--209:35, 2025.

\bibitem{DBLP:journals/twc/XieXZXGGP24}
G.~Xie, Z.~Xiong, X.~Zhang, R.~Xie, S.~Guo, M.~Guizani, and H.~V. Poor,
  ``{GAI-IoV}: Bridging generative {AI} and vehicular networks for ubiquitous
  edge intelligence,'' \emph{{IEEE} Trans. Wirel. Commun.}, vol.~23, no.~10,
  pp. 12\,799--12\,814, 2024.

\bibitem{DBLP:journals/iotj/ZhangSCCJ25}
M.~Zhang, X.~Shen, J.~Cao, Z.~Cui, and S.~Jiang, ``{EdgeShard}: Efficient {LLM}
  inference via collaborative edge computing,'' \emph{{IEEE} Internet Things
  J.}, vol.~12, no.~10, pp. 13\,119--13\,131, 2025.

\bibitem{DBLP:conf/lanman/MacarioSK25}
D.~Macario, H.~Seferoglu, and E.~Koyuncu, ``Model-distributed inference for
  large language models at the edge,'' in \emph{Proc. IEEE Int. Symp. Local
  Metropolitan Area Networks, \textnormal{Lille, France, July 7-8, 2025}}, pp.
  1--6.

\bibitem{10681712}
J.~Li, B.~Han, S.~Li, X.~Wang, and J.~Li, ``{CoLLM}: A collaborative {LLM}
  inference framework for resource-constrained devices,'' in \emph{Proc.
  IEEE/CIC Int. Conf. Commun. China, \textnormal{Hangzhou, China, Aug. 2024}},
  pp. 185--190.

\bibitem{DBLP:conf/icc/FengLLCZZTG25}
Z.~Feng, L.~Lu, Q.~Li, Y.~Chai, Z.~Zhang, Y.~Zhang, Y.~Teng, and D.~Guo,
  ``Distributed inference optimization for large language model in edge-cloud
  collaborative networks,'' in \emph{Proc. IEEE Int. Conf. Commun.,
  \textnormal{Montreal, QC, Canada, Jun. 8-12, 2025}}, pp. 6161--6166.

\bibitem{DBLP:journals/corr/abs-2502-12574}
C.~Luo, Z.~Cai, H.~Sun, J.~Xiao, B.~Yuan, W.~Xiao, J.~Hu, J.~Zhao, B.~Chen, and
  A.~Anandkumar, ``{HeadInfer}: Memory-efficient {LLM} inference by head-wise
  offloading,'' \emph{CoRR}, vol. abs/2502.12574, 2025.

\bibitem{DBLP:conf/icml/MaiYHYP24}
P.~Mai, R.~Yan, Z.~Huang, Y.~Yang, and Y.~Pang, ``Split-and-denoise: Protect
  large language model inference with local differential privacy,'' in
  \emph{Proc. Int. Conf. Mach. Learn., \textnormal{Vienna, Austria, July 21-27,
  2024}}.

\bibitem{DBLP:conf/mlsys/GimCLSK024}
I.~Gim, G.~Chen, S.~Lee, N.~Sarda, A.~Khandelwal, and L.~Zhong, ``Prompt cache:
  Modular attention reuse for low-latency inference,'' in \emph{Proc. Annu.
  Conf. Mach. Learn. Syst., \textnormal{Santa Clara, CA, USA, May 13-16,
  2024}}.

\bibitem{DBLP:journals/corr/abs-2505-13345}
S.~Luo, P.~Li, J.~Peng, H.~Wang, Y.~Zhao, Y.~Cao, Y.~Cheng, and T.~Chen,
  ``Occult: Optimizing collaborative communication across experts for
  accelerated parallel moe training and inference,'' \emph{CoRR}, vol.
  abs/2505.13345, 2025.

\bibitem{DBLP:conf/usenix/LiJZ0X23}
J.~Li, Y.~Jiang, Y.~Zhu, C.~Wang, and H.~Xu, ``Accelerating distributed moe
  training and inference with {Lina},'' in \emph{Proc. USENIX Annu. Tech.
  Conf., \textnormal{Boston, MA, USA, Jul. 10-12, 2023}}, pp. 945--959.

\bibitem{DBLP:journals/corr/abs-2509-25041}
Y.~Han, L.~Pan, J.~Peng, Z.~Tao, W.~Zhang, and Y.~Zhang, ``{GRACE-MoE}:
  Grouping and replication with locality-aware routing for efficient
  distributed moe inference,'' \emph{CoRR}, vol. abs/2509.25041, 2025.

\bibitem{DBLP:journals/corr/abs-2508-12851}
T.~Wu, L.~Wang, Z.~Wen, X.~Zhang, J.~Duan, X.~Zhang, and J.~Zuo, ``Accelerating
  edge inference for distributed moe models with latency-optimized expert
  placement,'' \emph{CoRR}, vol. abs/2508.12851, 2025.

\bibitem{DBLP:journals/corr/abs-2508-09208}
M.~Li, N.~Li, X.~Yuan, W.~Xu, Q.~Chen, S.~Guo, and H.~Zhang, ``{CoMoE}:
  Collaborative optimization of expert aggregation and offloading for moe-based
  {LLMs} at edge,'' \emph{CoRR}, vol. abs/2508.09208, 2025.

\bibitem{DBLP:journals/corr/abs-2503-01704}
M.~Hosseinzadeh and H.~Khamfroush, ``{DILEMMA:} joint {LLM} quantization and
  distributed {LLM} inference over edge computing systems,'' \emph{CoRR}, vol.
  abs/2503.01704, 2025.

\bibitem{DBLP:conf/mlsys/0002L0XV24}
S.~Bian, D.~Li, H.~Wang, E.~P. Xing, and S.~Venkataraman, ``Does compressing
  activations help model parallel training?'' in \emph{Proc. Annu. Conf. Mach.
  Learn. Syst., \textnormal{Santa Clara, CA, USA, May 13-16, 2024}}.

\bibitem{DBLP:journals/corr/abs-2412-04964}
Q.~Li, B.~Zhang, L.~Ye, Y.~Zhang, W.~Wu, Y.~Sun, L.~Ma, and Y.~Xie, ``Flash
  communication: Reducing tensor parallelization bottleneck for fast large
  language model inference,'' \emph{CoRR}, vol. abs/2412.04964, 2024.

\bibitem{DBLP:journals/corr/abs-2411-07942}
H.~Dong, T.~Johnson, M.~Cho, and E.~Soroush, ``Towards low-bit communication
  for tensor parallel {LLM} inference,'' \emph{CoRR}, vol. abs/2411.07942,
  2024.

\bibitem{DBLP:journals/corr/abs-2411-09510}
J.~Hansen{-}Palmus, M.~T. Le, O.~Hausd{\"{o}}rfer, and A.~Verma,
  ``Communication compression for tensor parallel {LLM} inference,''
  \emph{CoRR}, vol. abs/2411.09510, 2024.

\bibitem{DBLP:journals/corr/abs-2411-02829}
H.~Jin and Y.~Wu, ``Ce-collm: Efficient and adaptive large language models
  through cloud-edge collaboration,'' \emph{CoRR}, vol. abs/2411.02829, 2024.

\bibitem{DBLP:journals/corr/abs-2503-14882}
K.~Zhang, H.~He, S.~Song, J.~Zhang, and K.~B. Letaief,
  ``Communication-efficient distributed on-device {LLM} inference over wireless
  networks,'' \emph{CoRR}, vol. abs/2503.14882, 2025.

\bibitem{DBLP:conf/wiopt/KafetzisKK25}
D.~Kafetzis, R.~Khalili, and I.~Koutsopoulos, ``Large language model
  partitioning for low-latency inference at the edge,'' in \emph{Proc. Int.
  Symp. Modeling Optimization Mobile Ad Hoc Wireless Networks (WiOpt),
  \textnormal{Link{\"o}ping, Sweden, May 26-29, 2025}}, pp. 1--8.

\bibitem{DBLP:journals/corr/abs-2501-14205}
M.~Xu, D.~Niyato, and C.~G. Brinton, ``Serving long-context {LLMs} at the
  mobile edge: Test-time reinforcement learning-based model caching and
  inference offloading,'' \emph{CoRR}, vol. abs/2501.14205, 2025.

\bibitem{zhao-etal-2024-lingualinked}
J.~Zhao, Y.~Song, S.~Liu, I.~G. Harris, and S.~A. Jyothi, ``{LinguaLinked}:
  Distributed large language model inference on mobile devices,'' in
  \emph{Proc. Annu. Meet. Assoc. Comput. Linguist., \textnormal{Bangkok,
  Thailand, Aug. 2024}}, pp. 160--171.

\bibitem{DBLP:journals/corr/abs-2507-21276}
Y.~Li, Z.~Li, Y.~Zhu, and C.~Liu, ``{LeMix}: Unified scheduling for {LLM}
  training and inference on multi-{GPU} systems,'' \emph{CoRR}, vol.
  abs/2507.21276, 2025.

\bibitem{DBLP:conf/iwqos/ZhuZXD25}
J.~Zhu, L.~Zhao, F.~Xiao, and L.~Duan, ``Birds in cages: Edge inference
  allocation for distributed {LLM} deployment,'' in \emph{Proc. IEEE/ACM Int.
  Symp. Quality Service, \textnormal{Gold Coast, Australia, Jul. 2-4, 2025}},
  pp. 1--6.

\bibitem{10_1145/3768165}
F.~Wang, Z.~Zhang, X.~Zhang, Z.~Wu, T.~Mo, Q.~Lu, W.~Wang, R.~Li, J.~Xu,
  X.~Tang, Q.~He, Y.~Ma, M.~Huang, and S.~Wang, ``A comprehensive survey of
  small language models in the era of large language models: Techniques,
  enhancements, applications, collaboration with {LLMs}, and trustworthiness,''
  \emph{ACM Trans. Intell. Syst. Technol.}, 2025.

\bibitem{DBLP:conf/nips/VaswaniSPUJGKP17}
A.~Vaswani, N.~Shazeer, N.~Parmar, J.~Uszkoreit, L.~Jones, A.~N. Gomez,
  L.~Kaiser, and I.~Polosukhin, ``Attention is all you need,'' in \emph{Adv.
  Neural Inf. Process. Syst., \textnormal{Long Beach, CA, USA, Dec. 4-9,
  2017}}, pp. 5998--6008.

\bibitem{DBLP:conf/aaai/TangZWLXZ22}
C.~Tang, Y.~Zhao, G.~Wang, C.~Luo, W.~Xie, and W.~Zeng, ``Sparse {MLP} for
  image recognition: Is self-attention really necessary?'' in \emph{Proc. AAAI
  Conf. Artif. Intell., \textnormal{Virtual Event, Feb. 22-Mar. 1, 2022}}, pp.
  2344--2351.

\bibitem{DBLP:conf/nips/ZaheerGDAAOPRWY20}
M.~Zaheer, G.~Guruganesh, K.~A. Dubey, J.~Ainslie, C.~Alberti,
  S.~Onta{\~{n}}{\'{o}}n, P.~Pham, A.~Ravula, Q.~Wang, L.~Yang, and A.~Ahmed,
  ``Big bird: Transformers for longer sequences,'' in \emph{Adv. Neural Inf.
  Process. Syst., \textnormal{Virtual Event, Dec. 6-12, 2020}}.

\bibitem{DBLP:conf/iclr/XiaoTCHL24}
G.~Xiao, Y.~Tian, B.~Chen, S.~Han, and M.~Lewis, ``Efficient streaming language
  models with attention sinks,'' in \emph{Proc. Int. Conf. Learn. Represent.,
  \textnormal{Vienna, Austria, May 7-11, 2024}}.

\bibitem{DBLP:journals/tii/WangWJY25}
A.~Wang, G.~Wang, J.~Jiao, and S.~Yin, ``Self-attention sliding window enhanced
  canonical correlation analysis for incipient fault detection in dynamic
  industrial processes,'' \emph{{IEEE} Trans. Ind. Informatics}, vol.~21,
  no.~10, pp. 7412--7423, 2025.

\bibitem{DBLP:conf/naacl/HanWPX0JW24}
C.~Han, Q.~Wang, H.~Peng, W.~Xiong, Y.~Chen, H.~Ji, and S.~Wang,
  ``{LM-Infinite}: Zero-shot extreme length generalization for large language
  models,'' in \emph{Proc. Conf. North Am. Chapter Assoc. Comput. Linguist.:
  Hum. Lang. Technol., \textnormal{Mexico City, Mexico, Jun. 16-21, 2024}}, pp.
  3991--4008.

\bibitem{DBLP:conf/nips/JiangLZWLAHA0L024}
H.~Jiang, Y.~Li, C.~Zhang, Q.~Wu, X.~Luo, S.~Ahn, Z.~Han, A.~H. Abdi, D.~Li,
  C.~Lin, Y.~Yang, and L.~Qiu, ``{MInference 1.0}: Accelerating pre-filling for
  long-context {LLMs} via dynamic sparse attention,'' in \emph{Adv. Neural Inf.
  Process. Syst., \textnormal{Vancouver, BC, Canada, Dec. 10-15, 2024}}.

\bibitem{DBLP:journals/corr/abs-2110-11299}
L.~Liu, Z.~Qu, Z.~Chen, Y.~Ding, and Y.~Xie, ``Transformer acceleration with
  dynamic sparse attention,'' \emph{CoRR}, vol. abs/2110.11299, 2021.

\bibitem{DBLP:conf/icml/DasoulasSV21}
G.~Dasoulas, K.~Scaman, and A.~Virmaux, ``Lipschitz normalization for
  self-attention layers with application to graph neural networks,'' in
  \emph{Proc. Int. Conf. Mach. Learn., \textnormal{Virtual Event, Jul. 2021}},
  vol. 139, pp. 2456--2466.

\end{thebibliography}

\clearpage

\appendices
\section{Proof of Theorem \ref{theorem_1}}

For ease of exposition, we define the local and global self-attention operators by\begin{align}
	\mathcal{A}^m_n(\boldsymbol{X})= &\text{Attn}\left(\mathcal{P}^m_\text{Q}\left(\boldsymbol{\Pi}_n\boldsymbol{X}\right)\Big|\mathcal{P}^m_\text{K}\left(\boldsymbol{\Pi}_n\boldsymbol{X}\right),\mathcal{P}^m_\text{V}\left(\boldsymbol{\Pi}_n\boldsymbol{X}\right)\right),
\end{align}and\begin{align}
	&\check{\mathcal{A}}^m(\boldsymbol{X})= \text{Attn}\left(\mathcal{P}^m_\text{Q}\left(\boldsymbol{X}\right)\Big|\mathcal{P}^m_\text{K}\left(\boldsymbol{X}\right),\mathcal{P}^m_\text{V}\left(\boldsymbol{X}\right)\right).
\end{align}

We next introduce the following auxiliary notations. Let $\boldsymbol{X}^{h,t}$ and $\boldsymbol{O}^{h,t}$ denote the global hidden representations and attention output at the $h$-th local forward of the $t$-th communication round, i.e., $\boldsymbol{X}^{h,t}=\sum_{n=1}^{N}\boldsymbol{\Pi}_n \boldsymbol{x}_n^{h,t}$, and $\boldsymbol{O}^{h,t}=\sum_{n=1}^{N}\boldsymbol{\Pi}_n \boldsymbol{o}_n^{h,t}$. Let $\check{\boldsymbol{X}}^{h,t}$ denote the auxiliary hidden representations following the update dynamics of centralized self-attention. Specifically,  $\check{\boldsymbol{X}}^{h,t}$ is initialized from the global input embeddings as $\check{\boldsymbol{X}}^{1,0}=\boldsymbol{X}^\text{emb}$ and processed through Transformer blocks each with global self-attention computation given by\begin{align}\label{eq:update_local_x_cen}
	\check{\boldsymbol{O}}^{h,t}=	\check{\mathcal{A}}^m\left(\check{\boldsymbol{X}}^{h,t}\right),\end{align}followed by hidden representation refinement given by\begin{align}\check{\boldsymbol{X}}^{h+1,t}= \check{\boldsymbol{X}}^{h,t}+\check{\boldsymbol{O}}^{h,t}+\mathcal{F}^m\left( \check{\boldsymbol{X}}^{h,t}+\check{\boldsymbol{O}}^{h,t}\right),\end{align} with the final output denoted by $\boldsymbol{X}^{*}=\check{\boldsymbol{X}}^{1,T}$ upon completing all $M$ Transformer blocks. Note that $\check{\boldsymbol{X}}^{1,t+1}=\check{\boldsymbol{X}}^{H+1,t}$.

Before we show the main proof of \textbf{Theorem} \ref{theorem_1}, we first give \textbf{Theorem} \ref{theorem_2} below.
\begin{theorem}\label{theorem_2}Suppose \textbf{Assumptions} \ref{assumption1} and \ref{assumption2} hold. For any $h\in\{1,2,...,H-1\}$ and $t\in\mathcal{T}$, the deviation of hidden representations in each local forward between FedAttn and centralized attention follows that\begin{align}\label{theorem_2-1}&\left\Vert \boldsymbol{X}^{h+1,t}-\check{\boldsymbol{X}}^{h+1,t} \right\Vert_F\nonumber\\&\leq \left(1+\varrho_{Ht+h}\right)\left(1+\theta_{Ht+h}\right)\left\Vert\boldsymbol{X}^{h,t}-\check{\boldsymbol{X}}^{h,t}\right\Vert_F\nonumber\\&\quad+ \left(1+\theta_{Ht+h}\right)\sum_{n=1}^N\sigma_n^{Ht+h}.\end{align}\end{theorem}
\begin{IEEEproof}According to the update rule of hidden representations in (\ref{eq:update_local_x}) and (\ref{eq:update_local_x_cen}), it can be derived that\begin{align}&\left\Vert \boldsymbol{X}^{h+1,t}-\check{\boldsymbol{X}}^{h+1,t} \right\Vert_F\nonumber\\&=\bigg\Vert \boldsymbol{X}^{h,t}+\boldsymbol{O}^{h,t}+\mathcal{F}^m\left(\boldsymbol{X}^{h,t}+\boldsymbol{O}^{h,t}\right)\nonumber\\&\quad\quad-\left( \check{\boldsymbol{X}}^{h,t}+\check{\boldsymbol{O}}^{h,t}+\mathcal{F}^m\left( \check{\boldsymbol{X}}^{h,t}+\check{\boldsymbol{O}}^{h,t}\right)\right) \bigg\Vert_F\nonumber\\&=\bigg\Vert\boldsymbol{X}^{h,t}-\check{\boldsymbol{X}}^{h,t}+{\sum}_{n=1}^N\boldsymbol{\Pi}_n\mathcal{A}^m_n\left(\boldsymbol{X}^{h,t}\right)-\check{\mathcal{A}}^m\left(\check{\boldsymbol{X}}^{h,t}\right)\nonumber\\&\quad\quad+\mathcal{F}^m\nonumber\left(\boldsymbol{X}^{h,t}+{\sum}_{n=1}^N\boldsymbol{\Pi}_n\mathcal{A}^m_n\left(\boldsymbol{X}^{h,t}\right)\right)\nonumber\\&\quad\quad-\mathcal{F}^m\left( \check{\boldsymbol{X}}^{h,t}+\check{\mathcal{A}}^m\left(\check{\boldsymbol{X}}^{h,t}\right)\right)\bigg\Vert_F\nonumber\\&\leq\Big\Vert\boldsymbol{X}^{h,t}-\check{\boldsymbol{X}}^{h,t}\Big\Vert_F+\left\Vert\sum_{n=1}^N\boldsymbol{\Pi}_n\mathcal{A}^m_n\left(\boldsymbol{X}^{h,t}\right)-\check{\mathcal{A}}^m\left(\check{\boldsymbol{X}}^{h,t}\right)\right\Vert_F\nonumber\\&\quad\quad+\left\Vert \mathcal{F}^m\left(\boldsymbol{X}^{h,t}+{\sum}_{n=1}^N\boldsymbol{\Pi}_n\mathcal{A}^m_n\left(\boldsymbol{X}^{h,t}\right)\right)\right.\nonumber\\&\quad\quad-\mathcal{F}^m\left( \check{\boldsymbol{X}}^{h,t}+\check{\mathcal{A}}^m\left(\check{\boldsymbol{X}}^{h,t}\right)\right)\Big\Vert_F.\end{align}From (\ref{assumption1-2}), we have\begin{align}\label{proof-1}&\left\Vert \boldsymbol{X}^{h+1,t}-\check{\boldsymbol{X}}^{h+1,t} \right\Vert_F\nonumber\\&\leq\left\Vert\boldsymbol{X}^{h,t}-\check{\boldsymbol{X}}^{h,t}\right\Vert_F+\left\Vert\sum_{n=1}^N\boldsymbol{\Pi}_n\mathcal{A}^m_n\left(\boldsymbol{X}^{h,t}\right)-\check{\mathcal{A}}^m\left(\check{\boldsymbol{X}}^{h,t}\right)\right\Vert_F\nonumber\\&\quad+\theta_m\left\Vert\boldsymbol{X}^{h,t}-\check{\boldsymbol{X}}^{h,t}+\sum_{n=1}^N\boldsymbol{\Pi}_n\mathcal{A}^m_n\left(\boldsymbol{X}^{h,t}\right)-\check{\mathcal{A}}^m\left(\check{\boldsymbol{X}}^{h,t}\right)\right\Vert_F\nonumber\\&\leq\left(1+\theta_m\right)\left\Vert\boldsymbol{X}^{h,t}-\check{\boldsymbol{X}}^{h,t}\right\Vert_F\nonumber\\&\quad+\left(1+\theta_m\right)\left\Vert\sum_{n=1}^N\boldsymbol{\Pi}_n\mathcal{A}^m_n\left(\boldsymbol{X}^{h,t}\right)-\check{\mathcal{A}}^m\left(\check{\boldsymbol{X}}^{h,t}\right)\right\Vert_F\nonumber\\&=\left(1+\theta_m\right)\left\Vert\boldsymbol{X}^{h,t}-\check{\boldsymbol{X}}^{h,t}\right\Vert_F\nonumber\\&\quad+\left(1+\theta_m\right)\left\Vert{\sum}_{n=1}^N\boldsymbol{\Pi}_n\mathcal{A}^m_n\left(\boldsymbol{X}^{h,t}\right)-\check{\mathcal{A}}^m\left({\boldsymbol{X}}^{h,t}\right)\right.\nonumber\\&\quad+\check{\mathcal{A}}^m\left({\boldsymbol{X}}^{h,t}\right)-\check{\mathcal{A}}^m\left(\check{\boldsymbol{X}}^{h,t}\right)\Big\Vert_F\nonumber\\&\leq\left(1+\theta_m\right)\left\Vert\boldsymbol{X}^{h,t}-\check{\boldsymbol{X}}^{h,t}\right\Vert_F\nonumber\\&\quad+\left(1+\theta_m\right)\sum_{n=1}^N\boldsymbol{\Pi}_n\left\Vert\mathcal{A}^m_n\left(\boldsymbol{X}^{h,t}\right)-\boldsymbol{\Pi}_n^\top\check{\mathcal{A}}^m\left({\boldsymbol{X}}^{h,t}\right)\right\Vert_F\nonumber\\&\quad+\left(1+\theta_m\right)\left\Vert\check{\mathcal{A}}^m\left({\boldsymbol{X}}^{h,t}\right)-\check{\mathcal{A}}^m\left(\check{\boldsymbol{X}}^{h,t}\right)\right\Vert_F\nonumber\\&\leq\left(1+\varrho_m\right)\left(1+\theta_m\right)\left\Vert\boldsymbol{X}^{h,t}-\check{\boldsymbol{X}}^{h,t}\right\Vert_F\nonumber\\&\quad+\left(1+\theta_m\right){\sum}_{n=1}^N\sigma_n^m,\end{align}where the last inequality holds by (\ref{assumption1-1}) and (\ref{assumption2-1}). Substituting $m = Ht + h$ into (\ref{proof-1}), we obtain (\ref{theorem_2-1}). This concludes the proof of \textbf{Theorem} \ref{theorem_2}.\end{IEEEproof}
From \textbf{Theorem} \ref{theorem_2}, we obtain \textbf{Lemma} \ref{lemma_1} as follows.\begin{lemma}\label{lemma_1}Suppose \textbf{Assumptions} \ref{assumption1} and \ref{assumption2} hold. For any $t\in\mathcal{T}$, the deviation of hidden representations in each communication round between FedAttn and centralized attention follows that\begin{align}\label{fgdhfjsdfgf}&\left\Vert \boldsymbol{X}^{1,t+1}-\check{\boldsymbol{X}}^{1,t+1} \right\Vert_F\nonumber\\&\leq \left(\prod_{i=1}^{H} \Big(1+\theta_{Ht+i}\Big) \Big(1+\varrho_{Ht+i}\Big)\right)\left\Vert\boldsymbol{X}^{1,t}-\check{\boldsymbol{X}}^{1,t}\right\Vert_F\nonumber\\&\quad+\sum_{h=1}^{H-1} \left(\prod_{i=h+1}^{H} \left(1+\theta_{Ht+i}\right) \left(1+\varrho_{Ht+i}\right)\right)\nonumber\\&\quad\times\left(\left(1+\theta_{Ht+h}\right) \sum_{n=1}^N \sigma_{{Ht+h},n}\right).\end{align}\end{lemma}\begin{IEEEproof}From (\ref{theorem_2-1}), it can be derived that\begin{align}\label{dfsdg}&\left\Vert \boldsymbol{X}^{H,t}-\check{\boldsymbol{X}}^{H,t} \right\Vert_F\nonumber\\&\leq \left(\prod_{i=1}^{H-1} \Big(1+\theta_{Ht+i}\Big) \Big(1+\varrho_{Ht+i}\Big)\right)\left\Vert\boldsymbol{X}^{1,t}-\check{\boldsymbol{X}}^{1,t}\right\Vert_F\nonumber\\&\quad+\sum_{h=1}^{H-1} \left(\prod_{i=h+1}^{H-1} \left(1+\theta_{Ht+i}\right) \left(1+\varrho_{Ht+i}\right)\right)\nonumber\\&\quad\times\left(\left(1+\theta_{Ht+h}\right) \sum_{n=1}^N \sigma_{{Ht+h},n}\right).\end{align}From the update rule in the $H$-th local forward of each communication round as expressed in (\ref{eq:update_local_o_H}), we derive that\begin{align}\label{dsfd}&\left\Vert \boldsymbol{X}^{1,t+1}-\check{\boldsymbol{X}}^{1,t+1} \right\Vert_F\nonumber\\&=\left\Vert \boldsymbol{X}^{H,t}+\boldsymbol{O}^{H,t}+\mathcal{F}^{Ht+H}\left(\boldsymbol{X}^{H,t}+\boldsymbol{O}^{H,t}\right)\right.\nonumber\\&\quad\quad-\left(\check{\boldsymbol{X}}^{H,t}+\check{\boldsymbol{O}}^{H,t}+\mathcal{F}^{Ht+H}\left(\check{\boldsymbol{X}}^{H,t}+\check{\boldsymbol{O}}^{H,t}\right)\right) \Big\Vert_F\nonumber\\&\leq\left\Vert \boldsymbol{X}^{H,t}-\check{\boldsymbol{X}}^{H,t}\right\Vert_F\nonumber\\&\quad+\left\Vert\check{\mathcal{A}}^{Ht+H}\left(\boldsymbol{X}^{H,t}\right)-\check{\mathcal{A}}^{Ht+H}\left(\check{\boldsymbol{X}}^{H,t}\right)\right\Vert_F\nonumber\\&\quad+\left\Vert\mathcal{F}^{Ht+H}\left(\boldsymbol{X}^{H,t}+\check{\mathcal{A}}^{Ht+H}\left(\boldsymbol{X}^{H,t}\right)\right)\right.\nonumber\\&\quad\quad-\mathcal{F}^{Ht+H}\left(\check{\boldsymbol{X}}^{H,t}+\check{\mathcal{A}}^{Ht+H}\left(\check{\boldsymbol{X}}^{H,t}\right)\right) \Big\Vert_F\nonumber\\&\leq\left(1+\varrho_{Ht+H}\right)\left(1+\theta_{Ht+H}\right)\left\Vert \boldsymbol{X}^{H,t}-\check{\boldsymbol{X}}^{H,t}\right\Vert_F.\end{align}Substituting (\ref{dfsdg}) into (\ref{dsfd}), we obtain (\ref{fgdhfjsdfgf}). This concludes the proof of \textbf{Lemma} \ref{lemma_1}.\end{IEEEproof}From \textbf{Lemma} \ref{lemma_1}, we can derive that
\begin{align}\label{adefgbfds}&\Big\Vert \boldsymbol{X}^{1,T}-\check{\boldsymbol{X}}^{1,T}\Big\Vert\nonumber\\&\leq \sum_{t=0}^{T-1} \sum_{h=1}^{H-1} \left(\prod_{i=h+1}^{H} \left(1+\theta_{Ht+i}\right) \left(1+\varrho_{Ht+i}\right)\right)\nonumber\\&\quad\times\left(\left(1+\theta_{Ht+h}\right) \sum_{n=1}^N \sigma_{{Ht+h},n}\right) \left(\prod_{j=t+1}^{T-1} \left(\prod_{i=1}^{H} \Big(1+\theta_{Hj+i}\Big) \right.\right.\nonumber\\&\quad\times\Big(1+\varrho_{Hj+i}\Big)\bigg)\Bigg)+\prod_{t=0}^{T-1}\left(\prod_{i=1}^{H} \Big(1+\theta_{Ht+i}\Big) \Big(1+\varrho_{Ht+i}\Big)\right)\nonumber\\&\quad\times\left\Vert\boldsymbol{X}^{1,0}-\check{\boldsymbol{X}}^{1,0}\right\Vert_F.
\end{align}Note that $\boldsymbol{X}^{1,0}$ and $\check{\boldsymbol{X}}^{1,0}$ are both initialized from the global input embeddings $\boldsymbol{X}^\text{emb}$. Substituting $\left\Vert\boldsymbol{X}^{1,0}-\check{\boldsymbol{X}}^{1,0}\right\Vert_F=0$ into (\ref{adefgbfds}), we obtain (\ref{eafsgrthy}). This concludes the proof of \textbf{Theorem} \ref{theorem_1}.

\section{Proof of Corollary \ref{lemma_4}} 
Applying the uniform bounds $\theta_m\leq \theta$, $\varrho_m\leq \varrho$, and $\sigma_n^m\leq\sigma_n$ to (\ref{eafsgrthy}) in Theorem \ref{theorem_1} yields that\begin{align}\label{adefgbfdssdfgrtrf}&\left\Vert \boldsymbol{X}^{T}-\boldsymbol{X}^*\right\Vert_F\nonumber\\&\leq\sum_{t=0}^{T-1}\sum_{h=1}^{H-1} \left(\left(1+\theta\right) \sum_{n=1}^N \sigma_n\right)\left(\prod_{i=h+1}^{H} \left(1+\theta\right) \left(1+\varrho\right)\right)\nonumber\\&\quad \times\left(\prod_{j=t+1}^{T-1} \left(\prod_{i=1}^{H} \left(1+\theta\right) \left(1+\varrho\right)\right)\right)\nonumber\\&=\left(\left(1+\theta\right) \sum_{n=1}^N \sigma_n\right)\sum_{h=1}^{H-1} \left(\left(1+\theta\right) \left(1+\varrho\right)\right)^{H-h}\nonumber\\&\quad \times\sum_{t=0}^{T-1}\left( \left(1+\theta\right) \left(1+\varrho\right)\right)^{H(T-1-t)}\nonumber\\&=\left(\left(1+\theta\right) \sum_{n=1}^N \sigma_n\right)\frac{\left(\left(1+\theta\right) \left(1+\varrho\right)\right)^{H}-\left(1+\theta\right) \left(1+\varrho\right)}{\left(1+\theta\right) \left(1+\varrho\right)-1}\nonumber\\&\quad\times\frac{\left( \left(1+\theta\right) \left(1+\varrho\right)\right)^{HT}-1}{\left( \left(1+\theta\right) \left(1+\varrho\right)\right)^{H}-1}.\end{align}Substituting $M=HT$ into (\ref{adefgbfdssdfgrtrf}), we obtain (\ref{eafsgrthrthy}). This concludes the proof of \textbf{Corollary} \ref{lemma_4}.

\section{Discussion of Diminishing gains in communication efficiency} 

The Taylor series expansion of term \textit{(e)} around $\gamma = 1 + \varepsilon$ with $\varepsilon \to 0$ gives\begin{align}
		1-\frac{\gamma-1}{\gamma^H-1} &= 1-\frac{\gamma-1}{H \varepsilon + \frac{H}{2}(H-1) \varepsilon^2 + \mathcal{O}\left(\varepsilon^3\right)} \nonumber\\&= 1 - \frac{1}{H} + \mathcal{O}\left(\gamma - 1\right),\end{align}which establishes that, for any fixed number of local forwards $H$, FedAttn's approximation error asymptotically approaches the limit $1 - \frac{1}{H}$ as $\gamma\to 1$. The marginal reduction in communication overhead when increasing $H$ to $H+1$ is\begin{align}\frac{1}{H} - \frac{1}{H+1} = \frac{1}{H(H+1)},\end{align}with the marginal increase in approximation error being of the same magnitude. This reveals that as $H$ increases, the marginal reduction in communication overhead and the marginal increase in approximation error both decay quadratically as $\mathcal{O}\left(\frac{1}{H^2}\right)$, while the asymptotic convergence exhibits a first-order rate of $\mathcal{O}\left(\frac{1}{H}\right)$ whereby the communication overhead converges to zero and the approximation error converges to $1$. Specifically, the first increase in the number of local forwards, i.e., from $H=1$ to $2$, reduces the communication overhead by $\frac{1}{2}$ while introducing an approximation error of $\frac{1}{2}$. Subsequent increases in $H$ exhibit diminishing returns: $H=2$ to $3$ reduces communication overhead by $\frac{1}{6}$ with approximation error reaching $\frac{2}{3}$, while $H=3$ to $4$ achieves only a $\frac{1}{12}$ reduction with error reaching $\frac{3}{4}$. This suggests that small $H$ values exhibit substantial marginal effects in both communication overhead and approximation error, rendering this regime a significant trade-off between FedAttn's efficacy and efficiency. Conversely, in the large $H$ regime, each additional local forward pass yields progressively diminishing gains in FedAttn's communication efficiency, while performance degradation increasingly intensifies due to accumulated approximation errors.

\section{Proof of Theorem \ref{theorem_3}} 
From \textbf{Lemma} \ref{lemma_1}, we can derive that for any $t\in\mathcal{T}$, and $H_t$, the deviation of hidden representations in each communication round between FedAttn and centralized attention follows\begin{align}\label{fgdhfdajsdfgf}&\left\Vert \boldsymbol{X}^{1,t+1}-\check{\boldsymbol{X}}^{1,t+1} \right\Vert_F\nonumber\\&\leq \left(\prod_{i=1}^{H_t} \Big(1+\theta_{H_tt+i}\Big) \Big(1+\varrho_{H_tt+i}\Big)\right)\left\Vert\boldsymbol{X}^{1,t}-\check{\boldsymbol{X}}^{1,t}\right\Vert_F\nonumber\\&\quad+\sum_{h=1}^{H_t-1} \left(\prod_{i=h+1}^{H_t} \left(1+\theta_{H_tt+i}\right) \left(1+\varrho_{H_tt+i}\right)\right)\nonumber\\&\quad\times\left(\left(1+\theta_{H_tt+h}\right) \sum_{n=1}^N \sigma_{{H_tt+h},n}\right).\end{align}Therefore, we have\begin{align}&\Big\Vert \boldsymbol{X}^{1,T}-\check{\boldsymbol{X}}^{1,T}\Big\Vert\nonumber\\&\leq \sum_{t=0}^{T-1} \sum_{h=1}^{H_t-1} \left(\prod_{i=h+1}^{H_t} \left(1+\theta_{H_tt+i}\right) \left(1+\varrho_{H_tt+i}\right)\right)\nonumber\\&\quad\times\left(\left(1+\theta_{H_tt+h}\right) \sum_{n=1}^N \sigma_{{H_tt+h},n}\right) \left(\prod_{j=t+1}^{T-1} \left(\prod_{i=1}^{H_t} \Big(1+\theta_{H_tj+i}\Big) \right.\right.\nonumber\\&\quad\times\Big(1+\varrho_{H_tj+i}\Big)\bigg)\Bigg)+\prod_{t=0}^{T-1}\left(\prod_{i=1}^{H_t} \Big(1+\theta_{H_tt+i}\Big) \Big(1+\varrho_{H_tt+i}\Big)\right)\nonumber\\&\quad\times\left\Vert\boldsymbol{X}^{1,0}-\check{\boldsymbol{X}}^{1,0}\right\Vert_F.
\end{align}Substituting $\left\Vert\boldsymbol{X}^{1,0}-\check{\boldsymbol{X}}^{1,0}\right\Vert_F=0$ into (\ref{fgdhfdajsdfgf}), we obtain (\ref{sefrdgthyfjfd}). This concludes the proof of \textbf{Theorem} \ref{theorem_3}.

\vfill

\end{document}